\documentstyle[12pt,psfig]{article}

\newtheorem{lemma}{Lemma}

\newtheorem{theor}{\large\bf Theorem}

\def\FF{\hbox to 8.33887pt{\rm I\hskip-1.8pt F}}
\def\NN{\hbox to 9.3111pt{\rm I\hskip-1.8pt N}}
\def\PP{\hbox to 8.61664pt{\rm I\hskip-1.8pt P}}
\def\QQ{\rlap {\raise 0.4ex \hbox{$\scriptscriptstyle |$}}
{\hskip -4.5pt Q}}
\def\RR{\hbox to 9.1722pt{\rm I\hskip-1.8pt R}}
\def\ZZ{\hbox to 8.2222pt{\rm Z\hskip-4pt \rm Z}}
\def\ZZZ{Z\!\!\!Z}

\renewcommand{\thesection}{\Roman{section}}

\newcommand{\resetsect}{\setcounter{section}{1}}

\newcommand{\resetequ}{\setcounter{equation}{0}}
\newcommand{\Az}{{\cal S}}           
\newcommand{\fr}{{\cal F}}           
\newcommand{\tree}{{\cal T}}           

\newcommand{\be}{\begin{equation}}
\newcommand{\ee}{\end{equation}}
\newcommand{\bqa}{\begin{eqnarray}}
\newcommand{\eqa}{\end{eqnarray}}
\newcommand{\ba}{\begin{array}}
\newcommand{\ea}{\end{array}}
\newcommand{\p}[1]{{\partial\over \partial{#1}}}
\newcommand{\no}{\nonumber}

\newcommand{\lp}{\left (}
\newcommand{\rp}{\right )}
\newcommand{\qed}{\hfill \rule {1ex}{1ex}}
\newcommand{\al}{\alpha}
\newcommand{\bt}{\beta}

\newcommand{\de}{\delta}
\newcommand{\vep}{\varepsilon}

\newcommand{\th}{\theta}

\newcommand{\la}{\lambda}
\newcommand{\ph}{\phi}

\newcommand{\Om}{\Omega}
\newcommand{\Si}{\Sigma}
\newcommand{\La}{\Lambda}
\newcommand{\Lazero}{{\Lambda_{0}}}
\newcommand{\Lainv}{{\Lambda^{-2}}}

\newcommand{\Ga}{\Gamma}
\newcommand{\De}{\Delta}

\newcommand{\bpsi}{\bar{\psi}}

\begin{document}

\centerline{\large \bf Interacting Fermi liquid}
\centerline{\large \bf in two dimensions
at finite temperature}  \centerline{\large \bf Part I: Convergent Attributions}
\vskip 2cm

\centerline{M. Disertori and V. Rivasseau}

\centerline{Centre de Physique Th{\'e}orique, CNRS UPR 14}
\centerline{Ecole Polytechnique}
\centerline{91128 Palaiseau Cedex, FRANCE}

\vskip 1cm
\medskip
\noindent{\bf Abstract}

Using the method of continuous renormalization group around the Fermi
surface, we prove that a two-dimensional interacting
system of Fermions at low temperature $T$
is a Fermi liquid in the domain $ \lambda  |\log T|\le c  $ where $c$ 
is some numerical 
constant. According to [S1], this means that it is
analytic in the coupling constant $\lambda$, 
and that the first and second derivatives
of the self energy obey uniform bounds in that range. 
This is also a step in the program of rigorous (non-perturbative) study
of the BCS phase transition for many Fermions systems; 
it proves in particular that in dimension two the transition temperature 
(if any) must be non-perturbative in the coupling constant. The proof is
organized into two parts: the present paper deals with the convergent
contributions, and a companion paper (Part II) 
deals with the renormalization of dangerous two point subgraphs 
and achieves the proof.

\section{Introduction}
\resetequ

Conducting electrons in a metal at low temperature are well described
by Fermi liquid theory. However we know that the Fermi liquid theory is
not valid down to 0 temperature. Indeed below the BCS critical temperature
the dressed electrons or holes which are the excitations of the Fermi liquid
bound into Cooper pairs and the metal becomes superconducting.

During the last ten years a program has been designed to investigate
rigorously this phenomenon by means of field theoretic methods
[BG][FT1-2][FMRT1-3][S2].
In particular the renormalization group of Wilson and followers
has been extended to models with surface singularities such as the Fermi
surface. The ultimate goal is to create a mathematically rigorous
theory of the BCS transition and of similar phenomena of solid state physics.
This is a long and difficult program which requires to glue together several
ingredients in particular renormalization group around the Fermi
surface and spontaneous symmetry breaking.

A more accessible task is to precise the mathematical status of Fermi liquid
theory itself. Fermi liquid theory
is not valid at zero temperature because of the BCS instability. Even
when the dominant electron interaction is repulsive, the Kohn-Luttinger
instabilities prevent the Fermi liquid theory to be generically valid
down to zero temperature. There are nevertheless two proposals for a
mathematically rigorous Fermi liquid theory:

- one can block the BCS and Kohn-Luttinger
instabilities by considering models in which the Fermi surface
is not invariant under $p \to -p$ [FKLT]. In two dimensions it is possible to
prove (even non perturbatively) that in this case the Fermi liquid theory
remains valid at zero temperature, and the corresponding program is well under
way [FKLT].
However this program requires to control rigorously the stability of
a non-spherical Fermi surface under the renormalization group flow,
a difficult technical issue [FST];

- one can study the Fermi liquid theory at finite temperature
above the BCS transition temperature. A system of weakly interacting
electrons has an obviously stable thermodynamic limit
at high enough temperature, since 
the temperature acts as an infrared cutoff on the propagator
in the field theory description of the model. In this point of view,
advocated by [S1], the non trivial theorem consists in showing
that stability (i.e. summability of perturbation theory) holds
for all temperatures higher than a certain critical temperature
whose dependence in terms of the initial
interaction should be as precise as possible, and that the first and
second derivatives of the self-energy obey some uniform bounds. These
bounds rule out in particular Luttinger liquid behavior;
they do not hold in dimension 1, where Luttinger-liquid has been
established rigorously [BGPS]-[BM]. 

It is this second program that we do here. We prove an upper
bound on any critical temperature for two dimensional systems of Fermions
which is {\it exponentially small} in the coupling constant, hence
invisible in perturbation theory, and we check the uniform 
derivative bounds on the self-energy in that domain.
Our analysis relies on a  renormalization group analysis 
around the Fermi surface. Renormalization group flows were studied
perturbatively in the context of a spherical Fermi surface
in [FT2]. A non perturbative study in 2 dimension was performed
in [FMRT1], but it was limited to so called ``completely convergent graphs''.
In this paper we rely heavily on the ideas introduced in [FMRT1], but we
extend them to include non perturbative renormalization of the two point 
functions which allow the rigorous exponentially small upper bound.
This extension is not trivial since renormalization in phase space
in this context is complicated by the need for anisotropic sectors.
Also we use (in contrast with [FMRT1])
a {\it continuous} renormalization group scheme around the Fermi surface
(an other idea advocated in [S1]).
This scheme has been tested first in the simpler case of  the Gross Neveu
model (a field theory where there is no Fermi surface) in [DR1].

The next natural step in this program is to add the computation of
coupling constants flows (i.e. renormalization of four point functions).
This should be a rather straightforward extension of the methods of this
paper. It would allow to compute the optimal expected value $c_{o}$  
of the constant
$c$ in our upper bound on the critical temperature of Fermions systems.
A more difficult step is to glue this analysis to a kind
of $1/N$ expansion and to a bosonic analysis to control the region
at distance $\De_{BCS}\simeq e^{-c_{o}/\lambda}$ of the Fermi surface
[FMRT2].
In two dimensions and finite temperature we cannot expect true 
symmetry breaking by the Mermin-Wagner theorem, but we can expect
a Kosterlitz-Thouless phase for a two dimensional bosonic
field in a rotation invariant effective potential.
Finally at 0 temperature we have effectively a three-dimensional theory
(two dimensions for space, one for imaginary time). Continuous symmetry breaking
can then occur, with the associated Goldstone boson. The last part
of the analysis consists therefore in the non-perturbative control
of the infrared divergences
associated to this Goldstone boson, using Ward identities
at the constructive level [FMRT3].

Our result has quite a long proof,
which we organized therefore in two main parts. 
In this paper we introduce the model and prove the analyticity 
of the ``convergent contributions'' to the vertex functions, hence we
reproduce the results of [FMRT1], but with the continuous renormalization 
group technique. In a companion paper [DR2] we consider the complete sum 
of all graphs, perform renormalization of the two point subgraphs and obtain
our main theorem, with the bounds on the derivatives of the self-energy
proved in a separate Appendix.

\section{Model and Notations}
\resetequ

The simplest free continuum model for interacting Fermions
is the isotropic jellium model with a continuous rotation invariant
ultraviolet cutoff.
This model is rotation invariant, a feature which simplifies
considerably the study of the renormalization group flows
after introducing the interaction. In particular it
has a spherical Fermi surface.
It is a realistic model for instance in solid state physics
in the limit of weak electrons densities (where the Fermi surface becomes
approximately spherical).

The simplest Fermion interaction perturbing this free model
is a local four body interaction.
This is a realistic interaction for instance in a solid where the dominant
interaction is not the Coulomb interaction
but the electron-phonon interaction. After integrating out
the phonons modes 
an effective four body interaction is obtained, which is not
strictly local due to the non local phonon propagator. However at long
distances it is well approximated by a local interaction.
\footnote{Interaction with non-local but well-decaying
kernels can be added without much cost to our analysis.}

We use the formalism of non-relativistic field theory
at imaginary (periodic) time of [FT1-2][BG] to describe the interacting 
fermions at finite temperature. Our model is therefore similar to the 
Gross-Neveu model, but with a different, not relativistic propagator\footnote{
However there are some important differences:

- in GN the infrared singularity lies at $k=0$.
Renormalization subtracts divergent functions at this point. In
the  Fermi liquid the singularity
lies on the surface $k_0=0$, $|\vec{k}|=1$, so renormalization is
more complicated;

- in GN a natural infrared cut-off is given by the mass, in Fermi liquid
it is given by the temperature;

- in GN we are interested in the ultraviolet limit,
the low energy (renormalized) parameters being kept fixed;
in the Fermi liquid we fix the ultraviolet cut-off
and we want to deduce the long range properties from
the microscopic theory.}.

\subsection{Propagator without ultraviolet cutoff}
Using the Matsubara formalism, the propagator at temperature $T$,
$C(x_0,\vec{x})$, is antiperiodic in the variable
$x_0$ with antiperiod ${1\over T}$. This means that the
Fourier transform defined by
\be
 \hat{C}(k)= \frac{1}{2}\int_{-{1\over T}}^{1\over T} dx_0
\int d^2x \; e^{-ikx}\; C(x)
\ee
is not zero only  for   discrete
values (called the Matsubara frequencies) :
\be
k_0 =   \frac{2n+1}{\beta} \pi \ , \quad n \in \ZZ \ , \label{discretized}
\ee
where $\beta=1/T$ (we take $ /\!\!\!{\rm h} =k =1$). Remark that only
odd frequencies appear, because of  antiperiodicity.

Our convention is that a three dimensional vector is denoted by
$x = (x_0, \vec{x})$
where $ \vec{x}$ is the two dimensional spatial component.
The scalar product is
defined as $kx := - k_0 x_0 + \vec{k}\vec{x}$. 
By some slight abuse of notations we may write either
$C(x-\bar{x})$ or $C(x,\bar{x})$, where the first point corresponds to the field
and the second one to the antifield (using translation invariance of the
corresponding kernel).

Actually  $\hat{C}(k)$  is obtained
from the real time propagator by changing $k_0$ in $i k_0$ and is equal to:
\be
\hat{C}_{ab} (k) = \de_{ab} \frac{1}{ik_0-e(\vec{k})},
\quad \quad e(\vec{k})= \frac{\vec{k}^2}{2m}-\mu \ ,
\label{prop}
\ee
where $a,b \in \{1,2\}$ are the
spin indices. The vector $\vec k$ is two-dimensional. Since our theory
has two spatial dimensions and one time dimension,
there are really three dimensions.
The parameters $m$ and $\mu$ correspond to the effective mass and to
the chemical potential (which fixes the Fermi energy).
To simplify notation we put $2m= \mu=1$, so that
$e(\vec{k})= \vec{k}^2-1$.
Hence,
\be
C_{ab}(x) =\frac{1}{(2\pi)^2\beta}\; \sum_{k_0} \; \int d^2k\; e^{ikx}\;
\hat{C}_{ab}(k) \ .
\label{tfprop}\ee

The notation $\sum_{k_0}$ means really the discrete sum over the integer
$n$ in (\ref{discretized}).
When $T \to 0$ (which means $\beta\to \infty$) $k_0$
becomes a continuous variable, the corresponding discrete sum becomes an
integral, and the corresponding propagator
 $C_{0}(x)$ becomes singular
on the Fermi surface
defined by $k_0=0$ and $|\vec{k}|=1$.
In the following to simplify notations we will write:
\be
\int d^3k \; \equiv \; {1\over \beta} \sum_{k_0} \int d^2k
\quad , \quad 
\int d^3x \; \equiv \; {1\over 2}
\int_{-\beta}^{\beta}dx_0 \int d^2x \ . \label{convention}
\ee

In determining the spatial decay we will need the following lemma

\begin{lemma}
The function $C$ defined in  (\ref{tfprop}) can also be written
as
\be
C(x) = f(x_0,\vec{x}) :=
\sum_{m\in \ZZ} (-1)^m \; C_0\lp x_0+{m\over T}, \vec{x}\rp  \ .
\label{copie1}\ee
where $C_0$ is the propagator at $T=0$.
\end{lemma}
\paragraph{Proof}
To prove this lemma we first prove that the function $f$ is 
antiperiodic on ${1\over T}$. Since $
\hat{f}(k) = \hat{C}(k) \ \forall k
$, the Lemma holds.
\qed

 In this paper we do not perform yet any 
renormalization, hence we
do not introduce any counterterm, and the 
interaction is simply:

\be
S_V = \frac{\la}{2} \int_V d^3x\; (\sum_a \bpsi\psi)^2
\ee
where  $V:= [-\beta,\beta]\times V'$ and $ V'$ is an auxiliary volume
cutoff
in two dimensional space, that will be soon sent to infinity.
Remark that in (\ref{discretized}) $|k_0|\geq \pi/\beta\neq 0$ 
hence the denominator in $C(k)$ can never be 0 at non zero temperature.
This is why the temperature provides a natural infrared cut-off.

\subsection{Propagator with an ultraviolet cutoff}

It is convenient to add a continuous ultraviolet cut-off
(at a fixed scale $\La_0$) to the propagator
(\ref{prop}) for two reasons: first because it makes its Fourier transformed
kernel in position space  well defined, and second because a non relativistic
theory does not make sense anyway at high energies. To preserve physical (or
Osterwalder-Schrader) positivity one should introduce this ultraviolet cutoff
only on spatial frequencies [FT2]. However for convenience
we introduce this cutoff both on spatial and on Matsubara frequencies
as in [FMRT1]; indeed
the Matsubara cutoff could be lifted with little additional work.

For technical reasons it is also convenient to
introduce, as in [DR1], an auxiliary infrared cut-off at scale
$\La$, whose variation controls the renormalization group
flow. At the end the limit $\La\to 0$
is taken (we recall that the true infrared cutoff is the temperature,
which is not taken to 0 in this paper).
The propagator (\ref{prop})
equipped with these two cutoffs is called
$C^{\La_{0}}_{\La}$. It is defined as:
\be
C^{\La_0}_\La(k) :=  C(k)\left .
\left [u(r/\La^2_0)-u(r/\La^2)\right]\right |_{r=k_0^2+e^2(\vec{k})}
\ee
where we fixed $\La_0=1$ (for simplicity), $0\le \La \le 1$ and
the compact support function $u(r)\in{\cal C}_{0}^\infty({\rm R})$
satisfies:
\be
u(r)= 0 \quad {\rm for} \ |r|>1/2 \ ; \ u(r) =1
\quad {\rm for} \   |r|<1/4  \ ; \
 \int u(r) dr = 3/4  \ . \label{gevrey}
\ee
For later calculations it is useful to choose $u$ to be  a
Gevrey function\footnote{A
function $f\in {\cal C}^\infty ({\rm R}^d)$ with compact support is in
the
Gevrey class of order $s$ if there exist two constants $A$ and $\mu$ such that
\be
\forall n\geq 0,\quad ||f^{(n)}||_{_1} \leq A \mu^{-n} \lp \frac{n}{e}\rp^{ns}
\ee
and its Fourier transform satisfies (see [G]):
\be
\forall k\in {\rm R}^d \quad |\hat{f}(k)| \leq A
e^{-s\left (\frac{\mu}{\sqrt{d}}|k|\right )^{1/s}}
\ee
}.
The propagator can be  parametrized as:
\be
C^{\La_0}_\La(k) = \int_{\La_0^{-2}}^{\La^{-2}} d\al\;  C_\al(k)
\ee
where
\be
C_\al(k) =  C(k)\; \eta [\al\; r]\big |_{r=k_0^2+e^2(\vec{k})}
\qquad \eta(\al\; r) = -r u'(\al\; r) \ .
\ee
As $u'(\al\; r)\neq 0$ only  for $r\simeq 1/\al$
the propagator
$C_\al(k)$ is non zero only for
$ 1/2\sqrt{\al}\le \sqrt{k_0^2+e^2(\vec{k})}\le  1/\sqrt{2\al}$,
hence for  momenta in the volume between two tori in $R^{3}$ centered
on the critical circle $|\vec k| =1, k_{0}=0$
(see Fig.\ref{fermisurfig}):

\begin{figure}
\centerline{\psfig{figure=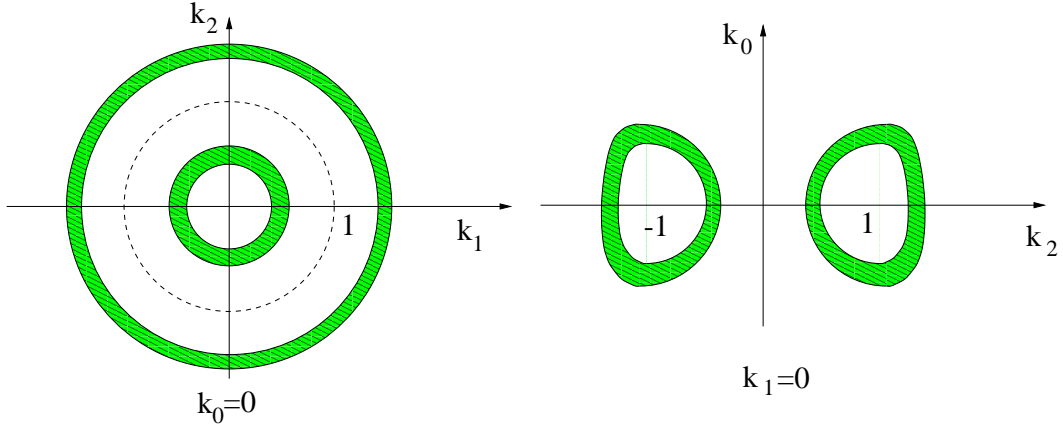,width=14cm}}
\caption{support of $C_\al$}
\label{fermisurfig}
\end{figure}

In short in the support of $C_\al(k)$ we have
$||\vec{k}|-1|\simeq \frac{1}{\sqrt\al}$ and $k_0\simeq 1/\sqrt\al$,
but they cannot be {\it simultaneously} much smaller.
Remark that the temperature cut-off implies that $C_\al=0$ if
$1/\sqrt{2\al}<\pi/\beta$, hence the real non zero propagator is
\bqa
C^{\La_0}_\La(k)& := &
\int_{\La_0^{-2}}^{\La_T^{-2}} d\al\;  C_\al(k)  \no\\
 &=& C(k)\left .
\left [u(r/\La^2_0)-u(r/\La_T^2)\right]\right |_{r=k_0^2+e^2(\vec{k})}
\label{propcutoff}\eqa
where we defined
\be
\La_T := \max \left [ \La\; ,\; \sqrt{2}\; \pi T \right ]\ . \label{propcutoff1}
\ee

\subsection{Vertex functions}

The vertex functions
are defined through the partition function:
\bqa
Z_V^{\La\Lazero}(\xi,\bar{\xi})& =& \int
d\mu_{C^{\Lazero}_\La}(\psi,\bpsi)
e^{-\Az_V(\psi,\bpsi)+<\psi,\xi >+ <\xi,\psi>}\no\\
<\psi,\xi > &=:& \int_{V} d^3x\; \bar{\psi}(x)\xi(x).
\eqa
where $\xi$ is an external field.
The $2p$-point vertex function is defined as:
\bqa
\Gamma^{\La\Lazero}(\{y\}, \{z\}) \hskip-.2cm &:=&\hskip-.2cm
\Gamma^{\La\Lazero}(y_1,...,y_p,z_1,...,z_p)\label{connected-function}\\
\hskip-.2cm &=& \hskip-.2cm  \lim_{V'\rightarrow \infty}
{\scriptstyle \de^{2p} \over
\scriptstyle\de\xi(z_1)..\de\xi(z_p)\de\bar{\xi}(y_1)..
\de\bar{\xi}(y_p)}
\left .\lp (\ln Z_V^{\La\Lazero} - F )(C^{\Lazero}_\La)^{-1}(\xi) \rp
\right |_{\xi=0}
\no\eqa
where $F (\xi)= <\xi, C^{\Lazero}_\La \xi>$ is the bare propagator.
These functions
are the coefficients of the effective action
(expanded in powers of the external fields)
at energy $\La$. They are in fact distributions (as easily seen because
there are graphs for which several external arguments hook to the same
vertex, hence create $\delta$ functions). Therefore we will later smear
the vertex functions $\Ga$ with smooth test functions $\ph_{1}(y_{1})$,...
$\ph_{p}(y_{p})$, $\ph_{p+1}(z_{1})$, ... $\ph_{2p}(z_{p})$ that
 are $L_\infty$ and $L_1$ in
position space. Actually, as we work at finite temperature, we can treat
test functions as propagators, that is introduce them 
at $T=0$ and then define the corresponding functions at $T\neq 0$.

Expanding the exponential in $Z$ we have:

\bqa
Z_V^{\La\Lazero}(\xi) &=&
\sum_{p=0}^\infty \frac{1}{p!^2} \sum_{n=0}^\infty
\frac{(-1)^{n}}{n!}    \la^{n}
\int_{V} d^3y_1...d^3y_p d^3z_1...d^3z_p d^3x_1...d^3x_n\no\\
&& \prod_{i=1}^p \xi(z_i)
\bar{\xi}(y_i)
 \left \{\ba{ccccccccccc}
y_{1} &...& y_{p}& x_{1}&
x_{1}&...&x_{n}&x_{n}\\
z_{1} &...& z_{p}& x_{1}&
x_{1}&...&x_{n}&x_{n}\label{blublu}\\
\ea
\right \}
\eqa
where we used Cayley's notation for determinants:
\be
\left \{\ba{c}
u_{i,a}\\ v_{j,b}\\ \ea \right \} =
\det (C^{\Lazero}_{\La, ab}(u_i-v_j)) \ .
\ee
The determinant is the sum over all Feynman graphs  amplitudes, and the
 logarithm selects the sum over connected graphs.  To obtain
$\log Z$ without expanding completely the determinant we use a forest formula.
Forest formulas are   Taylor expansions with integral
remainders which  test  links (here the propagators)
between $n\geq 1$ points (here the vertices)
and stop as soon as the final connected components are built.
The result is a sum over forests, a forest being a set of disjoint trees.

Like in [DR1] we use
the {\em ordered Brydges-Kennedy Taylor formula}, which states [AR1]
that for any smooth function $H$ of the $n(n-1)/2$ variables
$u_{l}$,  $l \in P_n = \{(i,j)| i,j\in \{1,..,n\}, i\neq j\}$,
\be
H |_{u_{l}=1} = \sum_{o-\fr} \lp \int_{0\le w_{1} \le ...\le w_k
\le 1}
\prod_{q=1 }^{k} dw_{q}
\rp
\lp \prod_{q=1 }^{k}\p{u_{l_{q}}}  H \rp ( w^{\fr}_{l}({w_{q}}), l \in P_n)
\label{bloblo}
\ee
where $o-\fr$ is any ordered forest, made of  $0\le k\le n-1$
links $l_{1},...,l_{k}$ over the $n$ points. To each link $l_{q}$  $q=1,...,k$
of $\fr$ is associated the parameter $w_{q}$, and to each pair $l=(i,j)$
is associated the weakening factor
$ w^{\fr}_{l}({w_{q}})$. These factors replace the variables
$u_{l}$ as arguments of the derived function $\prod_{q=1 }^{k}\p{u_{l_{q}}} H$
in (\ref{bloblo}).
These weakening factors $ w^{\fr}_{l}({w})$ are themselves functions
of the parameters $w_{q}$, $q=1,...,k$ through the formulas
\bqa
w^{\fr}_{i,i}(w)&=&1\no\\
w^{\fr}_{i,j}(w)&=&\inf_{l_{q}\in P^{\fr}_{i,j}}w_{q}, \quad\quad
\hbox{if $i$ and $j$ are
connected by $\fr$}\no\\&&
\hbox{where $P^{\fr}_{i,j}$ is the unique path in the forest
$\fr$ connecting $i$ to $j$}\no\\
w^{\fr}_{i,j}(w)&=&0 \quad \quad\hbox{if $i$ and $j$ are not
connected by $\fr$}.
\label{w-factor}\eqa
We apply this formula to the  determinant in (\ref{blublu}), inserting
the interpolation parameter $u_{l}$ in the UV cut-off $\La_0$ of
the covariance  $C^{\La(u)}_\La(x_i,x_j)$, when $i\neq j$. We define
$\La(u)$ by:
\be
\La^2(u)=  u (\La_0^2-\La^2) +\La^2
\ ; \  \La(0) =\La \;\;  ; \
\La(1) = \La_0 \ .
\ee

Now the product in (\ref{bloblo}) becomes:

\be
 \lp \prod_{q=1 }^{k}\p{u_{l_{q}}}  H \rp ( w^{\fr}_{l}({w_{q}}), l \in P_n)
=\lp \prod_{q=1 }^{k}\p{w_{q}} C^{\La(w_q)}_\La(k)(x_{l_q},y_{l_q})\rp
\det{\cal M}
\ee
which is the product of the forest line propagators and a remaining determinant
which contains all possible contractions of loop lines.
Actually, the elements of the matrix ${\cal M}$ are the loop line propagators
weakened by the forest formula.

Now, taking the logarithm of $Z$ and including (as announced above)
the smearing of external arguments by test functions we obtain a tree
expansion for the vertex function similar to the one of [DR1]:
\bqa
\Ga_{2p}^{\La\Lazero}(\phi_1,...\phi_{2p})\hspace{-.2cm}&=&\hspace{-.2cm}
\sum_{n=1}^\infty \frac{\la^n}{n!}
\sum_{o-\tree}\sum_{E}\sum_{\Om}  \vep(\tree, \Om)\int
d^3x_1...d^3x_{n}
\phi^{\La_T}_1(x_{i_{1}})...\phi^{\La_T}_{2p}(x_{j_{p}})\no \\
&&\int_{w_T\le w_{1} \le ...\le w_{n-1}\le 1}
\biggl[\prod_{q=1}^{n-1} \p{w_{q}}  C_{\La}^{\La(w_q)}
(x_{l_{q}}, {\bar x}_{l_{q}})dw_q \biggr]\det{\cal M}(E)\no\\
\label{sviluppo2}\eqa
where $o-\tree$ is the set of ordered trees over $n$ vertices,
  and $E$ is the set of pairs $(\phi_{j},v_{j})$ which
specifies which test function $\phi_{j}$ is hooked to which
internal vertex $v_{j}$ for $j=1,..., 2p$ (see [DR1]).
$\Om$
specifies for each tree line whether it comes from a $\psi\bpsi$ or
$\bpsi\psi$ contraction.  $\vep(\tree, \Om)$ is a global $\pm $ sign
whose exact (inessential) value is given in [AR2].
Finally
$w_T$ is defined by $\La(w_T)=\La_T$. Remark that
$w_T=0$ if $\La_T=\La$, i.e. if $\La \ge \sqrt{2}\pi T $,
and $w_T>0$ otherwise.
The bound $\La(w_i)\geq \La_T$ $\forall i$ is due to  
(\ref{propcutoff}-\ref{propcutoff1}).

In the following, as we are interested in the effective theory at
the energy $\La_T$,  we consider only external impulsions below
this energy. Therefore instead of $\phi$ we use
the test function with UV cut-off  $\phi^{\La_T}$ defined by
\be
\hat{\phi}^{\La_T}(k) :=
\hat{\phi}(k) \left. [u(r/\La^2_T)]\right |_{r=k_0^2+e^2(\vec{k})} .
\label{phiuv}\ee

\subsection{Bands}

The strategy to analyze (\ref{sviluppo2}) is similar to the one of
[DR1]. The determinant is bounded by
a Gram inequality (which gives no factorial)\footnote{The first example 
of combining a tree expansion with a Gram bound appears in [L].
We thank G. Gallavotti and
C. Wieczerkowski for pointing out this reference to us.}. Spatial integrals are
performed using the spatial decay of the tree
propagators $|\p{w}C^{\La(w)}_\La|$. To send the IR cut-off to zero
without generating unwanted factorials, we need to perform
some renormalization. These renormalizations, although more complicated than
in the field theory case, still involve
only two and four point subgraphs [FT1-2].
Therefore as in [DR1] we need to distinguish the so called dangerous subgraphs,
which means four-point and two-point quasi-local subgraphs. Remark that
a subgraph is called quasi-local if all internal lines have energy higher than
all external lines [R]. These contributions are decomposed into a
renormalized part with improved power counting, and a localized part
which in turn is absorbed into a flow of effective constants.

To implement this renormalization group program, the first tool is
to cut the momentum space into bands, which form a partition of unity.
The ordering of
the tree in the previous section cuts in a natural way
the  space of momenta into $n$ bands [DR1]. Indeed:
\be
w_T\leq w_1\leq w_2\leq...\leq w_{n-1}\leq 1 \rightarrow
\La_T \le \La (w_1)\leq\La (w_2)..\leq\La (w_{n-1})\leq \La_{0}\, .
\label{old-diff}\ee
The set of bands
is called $B=\{1,...,n\}$. The $q$-th band corresponds to
scales between $\La (w_{q-1})$ and   $\La (w_{q})$, where
we adopt the convention ${w}_{n}=1$
and  ${w}_0=w_T$ (hence $\La(w_0)=\La_T$ and
$\La(w_n)=\La_0$).
Then we can attribute each loop line to a well defined band.

On the other hand to external lines are associated the test functions
$\phi^{\La_T}_1(x_{i_{1}})$, ...$\phi^{\La_T}_{2p}(x_{j_{p}})$,
with a UV cutoff at $\La_T$, hence with impulsions lower than the
first band.
We can say that external lines belong to a first band with index 0, that
contains all impulsions at a distance at most $\La_T$ from
the Fermi surface.

Let's see how propagators for tree and loop lines look like.

\subsubsection{Tree propagators}

\begin{figure}
\centerline{\psfig{figure=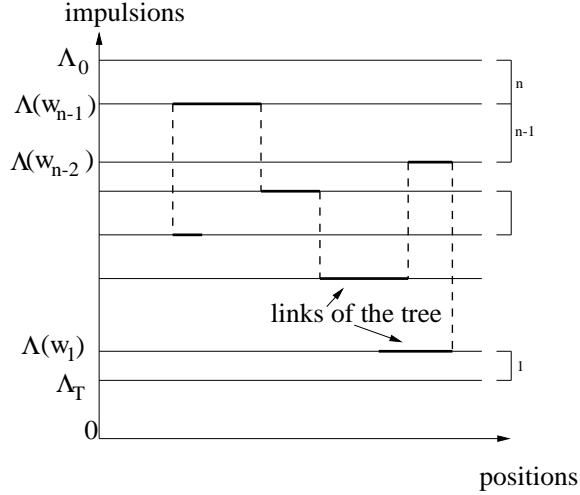,width=7.5cm}}
\caption{Band structure}
\label{bandfig}
\end{figure}

The $q$-th tree line propagator is given by:
\bqa
C^{w_{q}}(k)& =& \p{w_{q}}\int_{\La^{-2}(w_{q})}^{\La^{-2}} d\al\; C_\al(k)
=\frac{(\La_0^2-\La^2)}{\La^4(w_{q})}\; C_\al(k)\big |_{\al=
\La^{-2}(w_{q})}
\no\\
&=& \frac{(\La_0^2-\La^2)}{\La^4(w_{q})}\;
[ik_0 + e(\vec{k})]\; u'\lp \frac{k_0^2+ e^2(\vec{k})}
{\La^2(w_q)} \rp \ .
\label{treeline}\eqa
The derivative with respect to
$w_q$ fixes the $\al $ parameter of the line on the top of the band
$b_q$, and this tree line propagator
is considered by convention to belong to the $q$-th band.
In this way we have one tree line in each band, except the last one $b_n$ 
(see Fig.\ref{bandfig}).

\subsubsection{Loop lines}

Loop line propagators are the elements of the
$(n+1-p)\times(n+1-p)$ matrix $ {\cal M}(E)$:
\be
{\cal M}_{fg} =C^{\La(w^{\tree}_{f,g}(w))}_\La(x_f,x_g)\ .
\ee
The corresponding loop fields (respectively antifields) are labeled
by the index $f$ (respectively $g$).
Altogether they form a set $L$ labeled by an index
$a\in L=:\{1,...,2n+2-2p\}$, hence $a$ indexes both
the rows and columns of the determinant in  (\ref{sviluppo2}):
$a(f_1)=1,...,a(f_{n+1-p})=n+1-p,a(g_1)=n+2-p,...,a(g_{n+1-p})=2n+2-2p$.
Similarly to each tree line $l_{i}$ there corresponds two half tree lines
called $f_{i}$ and $g_{i}$. Each loop propagator can be
written as a sum of propagators restricted to single bands:
\be
C^{\La(w^{\tree}_{f,g}(w))}_\La(k) =
\sum_{j=1}^{i^{\tree}_{f,g} }
\int_{\La^{-2}(w_j)}^{\La^{-2}(w_{j-1})} d\al\;  C_\al(k)
 = C(k)  \sum_{j=1}^{i^{\tree}_{f,g} } u^j(k)
\label{band1}\ee
where   we define  $i^{\tree}_{f,g}$  as  the lowest index in the path
$P^{\tree}_{f,g}$ (defined in equation (\ref{w-factor}))
\bqa
i^{\tree}_{f,g}=   \inf\; \{ q \  | \  l_{q} \in P^{\tree}_{f,g}  \},
\eqa
and the function $u^j$ is the cutoff for the $j$-th band
\be
 u^j(k) :=
\left .\left [u[r\ \La^{-2}(w_j)]
- u[r\ \La^{-2}(w_{j-1})] \right ]\right |_{r=[k_0^2+e^2(\vec{k})] }\ .
\label{bco1}\ee

By multi-linearity one can expand the determinant in  (\ref{sviluppo2})
according to the different bands in the sum (\ref{band1}) for each
row and column:
\be
\det {\cal M}(E) =
\sum_\mu \det{\cal M}(\mu, E)\label{detmu}
\ee
where an attribution $\mu$ is a collection of band indices for each
loop field $a\in L$:
\be
 \mu= \{ \mu(f_1), ..\mu(f_{n+1-p}),\mu(g_1), ..\mu(g_{n+1-p})\} \ ,\
\mu(a)\in B  \ \ {\rm for}\   a=1...2n+2-2p.
\label{half-lines}
\ee

Now, for each attribution $\mu$ we need to exploit power counting. This
requires notations for the various types of fields or half-lines
which form the analogs of the quasi local subgraphs of [R] in our formalism
(that is subgraphs with all internal lines higher than the external ones).
For a loop half line (with index $a$) or an external line (with index $j$)
we call $v_{a}$ or $v_{j}$ the vertex to which it hooks.
Similarly, for tree half lines $f_{i}$ and $g_{i}$,
we call $v_{f_{i}}$ or $v_{g_{i}}$  the vertex to which they hook.
We define as $i_v$ the band index of the highest tree line hooked to the
vertex $v$, and, for each $k\geq 1$:
\be
T_k =  \{ l_{i} \in \tree |\; i \geq k\} \ .
\ee
In particular we define $t_k$  as the unique
connected component of $T_k$ containing the tree line $l_k$. We
say that a vertex $v\in t_k$ if $i_v\geq k$ and $l_{i_v}\in t_k$.
The matrix element of the determinant in (\ref{detmu}) is then
\bqa
{\cal M}_{fg}(\mu) (x_f,x_g)&=&
\de_{\mu(f),\mu(g)}\frac{1}{(2\pi)^2}
\int d^3k \; e^{ik(x_f-x_g)}C(k)\;u^{\mu(f)}(k)
 W^{\mu(f)}_{v_f,v_g}\; , \no\\
\eqa
where
\bqa
W^k_{v,v'}&=& 1 \quad {\rm if}\ v\ {\rm and}\ v' 
{\rm are\ connected\ by}\ T_k   \no\\
        &=& 0  \quad {\rm otherwise\ .}
\label{defw}\eqa
Now we define the quasi-local subgraph at level $k$  $g_k$ as
\bqa
g_k  &=& t_k \cup il_k\no\\
et_k &=& \{ l_i \in \tree | v_i\in t_k, i<k\}\no\\
il_k &=&\{ a\in L |v_a \in t_k, \mu(a) > {\cal A}(k) \} \no\\
el_k &=&\{ a\in L |v_a \in t_k, \mu(a)\leq {\cal A}(k) \}\no\\
ee_k &=&  \{ (\phi^{\La_T}_{j},v_{j}) \in E | v_j\in t_k \}\no\\
eg_k &=& et_k \cup el_k \cup ee_k\no\\
V_k &=& \{ v | v\in t_k \}
\label{definizioni}\eqa
where $et_k$, $el_k$ and $ee_k$ are the tree, loop and real
external external half lines respectively, and $il_k$ are the
internal loop half lines, and we
denoted by ${\cal A}(k)$ the index of the highest tree external line
of $g_k$.

In defining internal and external loop half-lines we
have observed that no new line connects to $t_k$ in the interval
between $k$ and ${\cal A}(k)$.
Hence all loop half-lines connected to the vertices of
$t_{k}$ with attributions between $k$ and
${\cal A}(k)$ are in fact internal
lines for the subgraph $g_k$ as they must contract between themselves.
Therefore we have considered as external loop half lines
only the ones with attributions $\mu(a)\leq {\cal A}(k)$.
In the following, we will  note by $|A|$ the number of elements in some set $A$.

\paragraph{Tadpoles}
Remark that $\mu(a)\leq i_{v_a}$ always. Indeed we could have
  $\mu(a)> i_{v_a}$  only if  the line $a$ belongs to a tadpole.
But the contribution of a tadpole is zero\footnote{Tadpoles are exactly zero
  because we choose our ultraviolet cutoff small enough. Otherwise
the tadpole would simply be very small, which would add some inessential
complications.}, as proved by the
following lemma:
\begin{lemma}
The amplitude of a tadpole with loop line in some band $i$  is zero
$\forall i$.
\end{lemma}
\paragraph{Proof}
The loop integral is:
\be
\frac{1}{(2\pi)^2}\int d^3 k\; C^{\La(w_i)}_{\La(w_{i-1})}(k) =
- \frac{1}{(2\pi)^2\beta}\sum_{ k_0}
\int d^2k \;\frac{ik_0+e(\vec{k})}{k_0^2+e^2(\vec{k})}
U\left [k_0^2, e^2(\vec{k})\right ]\ee
where
\be
U\left [k_0, e^2(\vec{k})\right ] =
\left [ u\lp \frac{k_0^2+e^2}{\La^2(w_i)}\rp -
u\lp \frac{k_0^2+e^2}{\La^2(w_{i-1})}\rp\right ].
\ee
 By the properties of $u$,
$U\neq 0$ only for $\La^2(w_{i-1})/4\leq k_0^2+e^2
\leq \La^2(w_i)/2$.
 The integral reduces to
\be
-  \frac{1}{(2\pi)^2\beta}\sum_{ k_0}
\int d^2k \frac{e(\vec{k})}{k_0^2+e^2(\vec{k})}
U\left [k_0^2, e^2(\vec{k})\right ]
\ee
as the other term in odd under $k_0$. Performing the change  of variables
 $t= |\vec{k}|^2-1$ the spatial integral (for any $k_0$ fixed)  becomes
\be
\int _0^{2\pi} d\th \int_{-1}^\infty \frac{dt}{2}
 \frac{t}{k_0^2+t^2}
U(k_0^2, t^2)=
\pi \int_{-1}^1 dt
 \frac{t}{k_0^2+t^2}
U(k_0^2, t^2)=0
\ee
by parity. Remark that the domain of $t$ can be reduced to
$[-1,1]$ since, for $t\geq 1$,
$k_0^2+t^2\geq 1> \La^2(w_i)/2$, hence $U=0$.
\qed

\subsubsection{Analyticity of convergent attributions}

We call an attribution $\mu$ convergent if it satisfies
$eg_k \ge 6$ for any $k>1$. Remark that for $k=1$,
$eg_1 = 2p$, and for $p\le 2$
we cannot require that this last subgraph has more than
4 external legs.

The convergent part of the theory is defined by the functions
\bqa
\lefteqn{\Ga_{2p,{\rm conv.}}^{\La\Lazero}
(\phi^{\La_T}_1,...\phi^{\La_T}_{2p})=}\no\\
&&\sum_{n=1}^\infty \frac{\la^n}{n!}
\sum_{o-\tree}\sum_{E\; \Om}  \vep(\tree, \Om)\int
d^3x_1...d^3x_{n}\;
\phi^{\La_T}_1(x_{i_{1}})...\phi^{\La_T}_{2p}(x_{j_{p}})\no \\
&&\int_{w_T\le w_{1} \le ...\le w_{n-1}\le 1}
\biggl[\prod_{q=1}^{n-1} \p{w_{q}}  C_{\La}^{\La(w_q)}
(x_{l_{q}},{\bar x}_{l_{q}})dw_q \biggr]\sum_{\mu \ {\rm conv.} }
\det{\cal M}(\mu, E)\ . \no\\
\label{sviluppoconv}\eqa

We start with a first theorem which essentially reproduces the result
of [FMRT1] in our framework of continuous cutoffs. This theorem states
that the infrared limit (i.e the zero temperature limit) of the convergent part
of the theory exists and is analytic in the bare coupling constant.

The full theorem on the Fermi liquid, which includes renormalization and
requires a finite temperature cutoff is postponed to the companion paper
(Part II).

\begin{theor}
For fixed  $\Lazero$ and $T\ge 0$, the limit $\La\to 0$
of the function \\
$\Ga_{2p,{\rm conv.}}^{\La\Lazero}(\phi^{\La_T}_1,...\phi^{\La_T}_{2p})$
exists and is analytic in $\la$ for any
$|\la|\leq c$ where $c$ is the convergence radius.
\end{theor}

This partial result is interesting because it isolates the constructive
arguments from the computation of the renormalization group flow.
We conjecture that the same theorem holds in three
dimensions but have no proof until now
(see however [MR] for a partial result in that direction). The rest
of the paper is devoted to a proof of Theorem 1. 

\section{Further Expansion Steps}
\resetequ
\subsection{Chains}

The decomposition into bands has a price, that is
we have to perform the additional sum over
convergent attributions $\mu$.
As in [DR1] this sum  might develop a factorial.
In other words  fixing the
band index for each single half-line
develops too much the determinant. To overcome this difficulty
we remark that the attributions contain much more
information than necessary, hence we can
group  attributions into packets to reduce the number of
determinants to bound. This operation is based on four remarks.
For each band index
 $i$  we analyze the subgraph $g_i$:
\begin{itemize}
\item{}
for each $g_i$ nothing happens   in the interval
between $i$ and ${\cal A}(i)$, as it contains just loop internal half lines
that contract between themselves.
Therefore we can regroup all the attributions in this interval;
\item{} if $|eg_i|\leq 10$
we want to know exactly which loop
fields are external and which ones are internal;
\item{} if $|eg_i|\geq 11$ and
$|et_i|+|ee_i|< 11$  we just want to fix
the attributions for
$11-|et_i|-|ee_i|$ loop fields,
but we do not need to fix the attributions for the
remaining loop fields;
\item{} if $|eg_i|\geq 11$ and
$|et_i|+|ee_i| \geq 11$  we do not fix
the attributions for  any loop field.
\end{itemize}
Remark that a subgraph is potentially divergent when it has two or
four external lines.
For this reason in [DR1] we selected at most five external lines
to ensure convergence. Here we select at most eleven external lines
because of additional technical difficulties due to the sector counting
and renormalization, that will be explained in the following. 
As seen below this does
not develop too much the determinant.

Hence, instead of expanding the loop determinant over lines and columns
as a sum over all attributions
\be
\det {\cal M} = \sum_{\mu} \det{\cal M}(\mu)
\ee
we write it as a sum over a smaller
set $\cal P$ (called the set of packets). These packets
are defined by means of a function
\bqa
\phi: \{\mu\} & \longrightarrow & {\cal P}\no\\
      \mu  &\mapsto & {\cal C} = \phi(\mu)
\eqa
which to each attribution $\mu$ associates
a class ${\cal C} = \phi(\mu)$ element of ${\cal P}$.
For our resummation purpose, the function $\phi$ must have two crucial
properties:
\begin{itemize}
\item{} $\#\{{\cal P}\}\leq K^n $ (this is critical for summation over packets);
\item{} there exists a matrix ${\cal M'}$ such that
\be
\sum_{\mu\in \phi^{-1}({\cal C})} \det{\cal M}(\mu)
= \det {\cal M'}({\cal C})
\ee
and some form of Gram's inequality applies to $\det {\cal M'}({\cal C})$.
\end{itemize}
The construction of a function $\phi$ with these properties is
developed in detail in [DR1] \footnote{We need only to modify $\phi$ slightly
to accommodate the expansion up to eleven  external lines 
instead of five external lines.
This has no other consequences than a larger constant $K$ for the
first condition (the number $3^{5}$ in [DR,(IV.13)] is replaced
by $3^{11}$).}. We just recall the result:
for each class ${\cal C}$, each loop field $a$ belongs
no longer to a single band $\mu(a)$, but
to a set of bands:
\be
J_a({\cal C})= \{\mu(a)| m(a,{\cal C})\leq \mu(a) \leq
M(a,{\cal C})\leq i_{v_a}\}
\ee
and the new matrix elements are
\be
{\cal M'}_{x_f,x_g}({\cal C})=\frac{1}{(2\pi)^2}
\int d^3k\; e^{ik(x_f-x_g)} C(k)
\sum_{q=1}^n \eta_{a(f)}^q\eta_{a(g)}^q u^q(k)
W^q_{v_f,v_g}
\label{newm}\ee
where ${\cal M}'$ is a function of ${\cal C}$, and
$\eta_a$ is the characteristic function of the set of bands
attributed by ${\cal C}$ to the loop
field $a$:
\be
\eta_a({\cal C}): B \rightarrow \{0,1\} \quad
\eta_a^q({\cal C}) =
\left\{\ba{cc}
0 & \hbox{if}\; q\not\in J_a({\cal C})\\
1 & \; \  \hbox{if}\; q\in J_a({\cal C})\ .\\
\ea\right.
\label{etaf}\ee

Finally we remark that the construction of [DR1] groups
convergent attributions $\mu$ into {\it convergent classes} ${\cal C}$
which form a subset of the set ${\cal P}$. Therefore the convergent
functions $\Ga_{2p,\; {\rm conv.}}^{\La\Lazero}
(\phi^{\La_T}_1,...\phi^{\La_T}_{2p})$
can be rewritten as:

\bqa
\lefteqn{\Ga_{2p,\; {\rm conv.}}^{\La\Lazero}
(\phi^{\La_T}_1,...\phi^{\La_T}_{2p})=}\no \\
&&\sum_{n=1}^\infty \frac{\la^n}{n!}
\sum_{o-\tree}\sum_{E}\sum_{\Om}\sum_{{\cal C}_{\rm c}}
\vep(\tree, \Om)\int
d^3x_1...d^3x_{n}
\phi^{\La_T}_1(x_{i_{1}})...\phi^{\La_T}_{2p}(x_{j_{p}})\no \\
&&\int_{w_T\le w_{1} \le ...\le w_{n-1}\le 1}
\biggl[\prod_{q=1}^{n-1}  C^{w_q}
({\bar x}_{l_{q}}, x_{l_{q}})dw_q \biggr]\det{\cal M}'({\cal C},E)
\label{conv}\eqa
and the definitions of  internal and external lines for each subgraph
$g_i$ can be generalized:
\bqa
il_i({\cal C})& :=& \{a\in L|v_a\in t_i, M(a,{\cal C}) > {\cal A}(i)\}
\no\\
el_{i}({\cal C})& :=& \{a\in L|v_a\in t_i,  M(a,{\cal C})\leq
{\cal A}(i)\}\no\\
eg_i({\cal C})& := & et_i\cup el_{i}({\cal C})\cup ee_{i}\ .
\label{conn1}\eqa

\subsection{Partial ordering}

We have seen that attributions contain much more information
than necessary and that this affects the convergence of the
series. Hence we have regrouped attributions into packets
preserving only the information  to perform power counting.

Similarly the total ordering over tree line
energies  contains unnecessary information that make power counting more
complicated and less transparent. Indeed we are not interested in 
the relative ordering of tree lines that belong to mutually
disjoint connected components $g_i$.
Hence we reorganize the scale analysis according to a structure
that we call  Clustering Tree Structure ($CTS$), that contains
the desired scale information and no more. This structure is
closely related to the ``Gallavotti-Nicolo'' trees.

\begin{figure}
\centerline{\psfig{figure=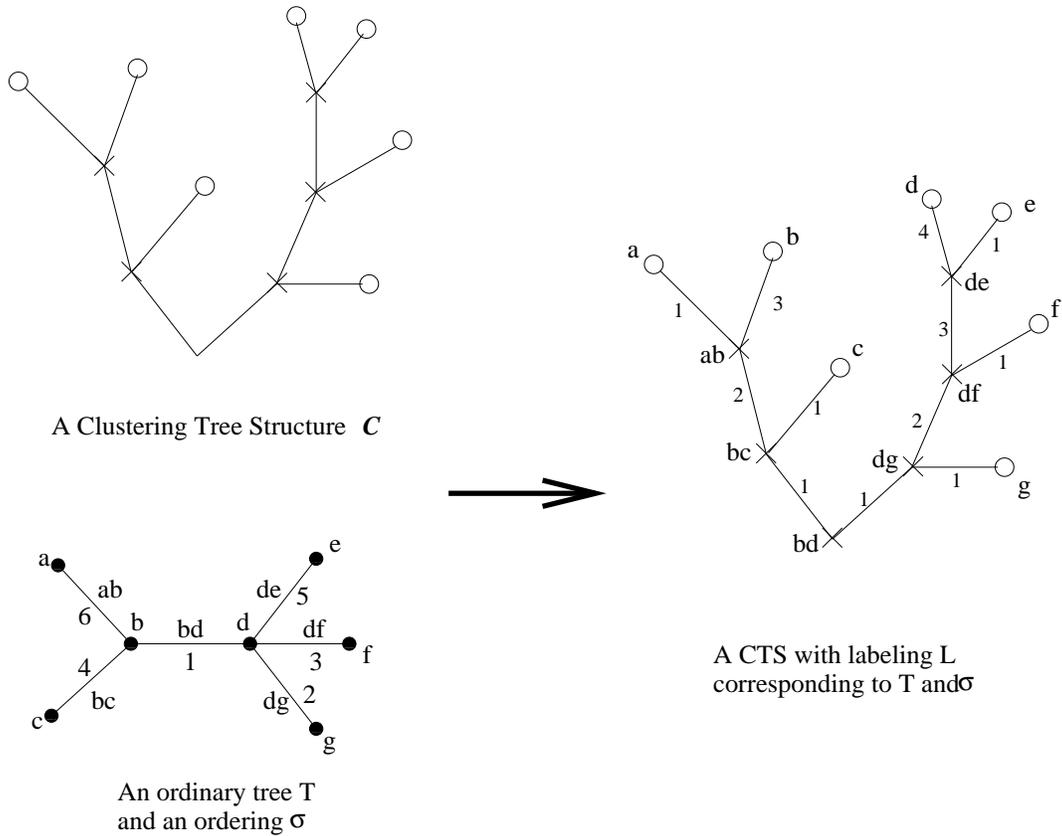,width=14cm}}
\caption{left: A $CTS$ and a tree, with an ordering; right: The associated CTS
with labeling induced. The vertices of the tree are named as a,b,c,d,e,f,g;
the lines are named by the pair of vertices they join; the ordering is
indicated by numbers 1,2,3,4,5,6 on the lines of $\tree $.
 Finally on the right,
the numbers $N_{\ell}(\tree ,{\cal L})$ are shown on each line $\ell$.}
\label{ctsfig}
\end{figure}

\paragraph{Definition} A clustering Tree Structure $CTS$ is
an unlabeled rooted tree, with $2n-2$
lines and $2n-1$ vertices of two different types~:
$n-1 $ crosses and $n$ dots, such that the root is a cross with
coordination 2, each other cross has coordination 3 and each dot
coordination 1 (see Fig.\ref{ctsfig}).

Obviously
\begin{lemma} The number of $CTS$ at order $n$ is at most $3^{n-1}$.
\end{lemma}

\noindent{\bf Proof}: We start from the cross root and climb in the structure.
At each cross there are at most three choices for the two vertices
immediately above~: two dots, one dot and one cross, or two crosses.
Hence the number of crosses being $n-1$ the total number of
choices is bounded by $3^{n-1}$ (this is only an upper bound because
some choices may not lead to a structure made of
$n-1 $ crosses and $n$ dots).

\qed

\subsubsection{Labeling}
We want to relate a $CTS$ at order $n$ to an ordinary tree $\tree$ with $n$
vertices. The $n-1$ lines of $\tree$ are labeled by an index $l$ and
the $2n-2$ lines of the $CTS$ are labeled by an index $\ell$ (they
should not be confused). A labeling ${\cal L}$ of the $CTS$ is a one to
one map between the set of vertices (crosses and dots) of the $CTS$ and
the  vertices and lines of $\tree$, so that each cross of the $CTS$
is labeled by a particular line of $\tree$, and each dot of the $CTS$ by a
particular
vertex of $\tree$, satisfying a further constraint. For each
$\ell$, let $T_{\ell}({\cal L})$ be the subset of $\tree$ made of all lines
and
vertices of $\tree$ corresponding to all crosses and dots ``above $\ell$''
(that is
such that the unique path in $CTS$ joining this cross or dot to the
root passes through $\ell$). The constraint on the labeling
${\cal L}$ is that $T_{\ell}({\cal L})$ has to be connected for all $\ell$.
We call $N_{\ell}(\tree,{\cal L})$ the number of external lines of
$T$ hooked to $T_{\ell}({\cal L})$.
\[
{\cal L } \{\times, \circ \}  \longrightarrow  \{l,v\}\quad
{\cal L}(\times) =  l\quad {\cal L}(\circ) = v\ .
\]
We consider only in what follows trees $\tree$ with coordination $N_{v}$ at
each vertex $v$ bounded by 4 (since other trees cannot appear as subgraphs
in the model we consider). Remark that a tree can be considered as
the list $V=\{N_{v}\}$ of its coordination numbers
plus the set of Wick contractions $W$ which
associates together two by two the half lines or ``fields'' hooked
to each vertex, subject to the constraint that the resulting
graph is a tree.

Let $\tree$ be a tree with $n$ vertices, and $\sigma_T$ a total ordering of
its lines. In [DR1] it is shown how to construct an associated $CTS$
and a labeling ${\cal L}$.
We recall the rule~: the first line in the ordering is the cross root.
When cut, it separates $\tree$ into two ordered trees $\tree_{1}$ and 
$\tree_{2}$
(possibly reduced to a single vertex). The process is iterated in each
subtree: in $\tree _{1}$ and $\tree _{2}$ 
the lowest lines give the label of the
crosses immediately above the root and so on (see Fig.\ref{ctsfig}).
When subtrees reduced to a single vertex are met, a dot
appears instead of a cross.

Conversely for a given tree $\tree$, the same $CTS$ and labeling ${\cal L}$
can be obtained from many total orderings  $\sigma_T$. Indeed $CTS$ and
${\cal L}$ induce only a {\it partial} ordering $\sigma_P $
on the lines of $\tree$~:
$l_{i} \ge_P l_{j}$ if the path from the cross
with label $l_{i}$ to the root
passes through the cross with label $l_{j}$.  Every total ordering
$\sigma_T$ compatible with this partial ordering gives
the same  $CTS$ and labeling ${\cal L}$. This is somehow a defect. Our new
point of view resums all these total orderings to retain
only the partial ordering $\sigma_P$
(which is the one relevant for scale analysis).

Hence the sum over ordered trees can be written as
\[
\sum_{o-\tree} = \sum_{u-\tree} \sum_{\sigma_T}
= \sum_{u-\tree} \sum_{CTS} \sum_{\cal L} 
\sum_{\sigma_T\rightarrow(CTS,{\cal L})}
=\sum_{CTS} \sum_{u-\tree}\sum_{\cal L} 
\sum_{\sigma_T\rightarrow(CTS,{\cal L})}
\]
where $u-\tree$ is an unordered tree and
$\sum_{\sigma_T\rightarrow(CTS,{\cal L})}$ is the sum over
the set of total orderings that give the same couple $(CTS,{\cal L})$,
for $u-\tree$ fixed.
Now we observe that
\[
\sum_{\sigma_T\rightarrow(CTS,{\cal L})}
\int_{w_T\le w_{1} \le ...\le w_{n-1}\le 1} =
\int_{w_T\le w_{{\cal A}(i)} \le w_{i}\le 1, \ \forall i}
\]
where the integration is now on the region of the $w$'s parameters 
satisfying the
partial ordering relations associated to $\sigma_P$. 
We call $w_r := \min_i w_i$ the parameter associated to the lowest
tree line, that is the root of the $CTS$, and by convention we put
$w_{{\cal A}(r)}:= w_T$.
Remark that now  for any $w_i$ we only know that
\be
\min [w_{i'},w_{i''}] \geq w_i \geq w_{{\cal A}(i)}
\label{bornew}\ee
where $w_{i'}$ and  $w_{i''}$ are the parameters associated to the
two crosses above $i$ (if there is a dot instead we assume $w_{i'}=1$).
In this new point of view the band $q$ corresponds to the energy interval
$[\La(w_q),\La(w_{{\cal A}(q)})]$ instead of 
$[\La(w_q),\La(w_{q-1})]$ and in (\ref{newm}), the new
matrix element $W^q_{v_f,v_q}$ selects only the vertices 
connected by $t_q$, hence in (\ref{defw}) $T_k$ has to be replaced by $t_{k}$. 
The expression (\ref{conv}) for the vertex function becomes
\bqa
\lefteqn{\Ga_{2p,\; {\rm conv.}}^{\La\Lazero}
(\phi^{\La_T}_1,...\phi^{\La_T}_{2p})=}
\no \\
&&\sum_{n=1}^\infty \frac{\la^n}{n!} \sum_{CTS}
\sum_{u-\tree} \sum_{\cal L}\sum_{E}\sum_{\Om}\sum_{{\cal C}_{\rm c}}
\vep(\tree, \Om)\int
d^3x_1...d^3x_{n}
\phi^{\La_T}_1(x_{i_{1}})...\phi^{\La_T}_{2p}(x_{j_{p}})\no \\
&&\int_{w_T\le w_{{\cal A}(i)} \le w_{i}\le 1}
\biggl[\prod_{q=1}^{n-1}  C^{w_q}
(x_{l_{q}},{\bar x}_{l_{q}})dw_q \biggr]\det{\cal M}'({\cal C},E)\ .
\label{convpo}\eqa

\subsection{Sectors}

Band decoupling is not enough to obtain correct power counting.
Roughly speaking, this happens for two reasons.

\noindent{\bf 1.}
The partition of unity for internal lines (tree and loop lines)
is not fine enough, as the volume in
phase space $\Delta x \Delta k$ depends on $\al$.
Actually $\Delta x$ is given by
the rate of spatial decay which is $1/\sqrt\al$ in all three directions.
On the other hand $\Delta k$ is given by the band volume,
proportional  to $1/\al$. Then
$\Delta x \Delta k \simeq \sqrt\al $. To obtain a
phase space volume independent from $\al$ we must take a smaller volume
in the momentum space. For that, adapting to our continuous
formalism the idea of [FMRT1], we cut the two
dimensional Fermi surface $|\vec{k}|=1$ into angular sectors of size
$1/\al_s^{1/4}$
\footnote{$\al_s$ is not necessarily equal to $\al$, since we need
to exploit momentum conservation of sectors at various intermediate scales
between $\al$ and the ultraviolet scale. The power
${1/4}$ is chosen as in [FMRT1], to avoid a logarithmic divergence
related to ``almost collapsed rhombuses''.}.
Now the volume in phase space of a single angular sector is
$1/(\al \al_s^{1/4})$. The spatial decay rate is $1/\sqrt{\al}$ on two
directions, and $1/\al_s^{1/4}$ on the third one, tangential
direction (provided $\al_s$ is not bigger than $\al$, as explained in [FMRT1]).
Then the phase space volume becomes a constant
independent from $\al$ and $\al_s$, as it should for
a single ``degree of freedom'' of the theory.

\noindent{\bf 2.}
When $2p>0$ we need to cut the support of $\hat{\phi}^{\La_T}$ into angular
sectors in order to exploit momentum conservation, at least for
subgraphs $g_i$  with $|eg_i({\cal C})|\leq 10$
(as in this case we know all the external lines of the subgraph).

\subsubsection{Sector Cutoffs}

To introduce the angular sectors we insert in
\[
C_\al(x) = {1 \over (2\pi)^{2}\beta}\sum_{k_0}\int_{0}^{\infty}
\  d|k|\  |k|
\ \int_0^{2\pi} d\theta \
e^{ik x}
C_\al(k)
\]
and in
\[
\phi^{\La_T}_i(x) =  {1 \over (2\pi)^{2}\beta}
\sum_{k_0}\int_0^\infty d|k|\; |k| \int_0^{2\pi} d\th \;
e^{ikx}  \;
u(r/\La_T)
\; \hat{\phi}_i(k)
\]
the unitary integral
\be
{4\over 3}\al_s^{1/4} \int_0^{2\pi} d\th_s \; \chi^{\th}_{\al_s}(\th_s) =1\ ,
\label{sec1}\ee
where
$ \chi^{\th}_{\al_s}(\th_s)= \chi^{\th_s}_{\al_s}(\th) $
selects a small angular
sector centered on $\th_s$. The factor  ${4\over 3}\al_s^{{1/4}}$
is needed to normalize properly the integral (see (\ref{gevrey})).
Indeed to define $\chi$ we use again the Gevrey function
$u:{\RR}\rightarrow {\RR}$ of the previous section:
\be
\chi^{\th}_{\al_{s}}(\th_s) := u_p^{\al_s}[\al_s^{1/4}(\th-\th_s)]\ ,
\ee
where $u_p^{\al_s}$ is the periodic function of period
$\tau=2\pi\al_s^{1/4},$ obtained from $u$ by:
$u_p^{\al_s}(y)=u(x)$ when $y= x+n\tau$ for some
$x\in [-1/2,1/2[$ and $n\in  {\ZZZ}$,
and $u_p^{\al_s}(y)=0$ otherwise.
This definition satisfies the condition (\ref{sec1}).

A {\it sector}  is {\it defined} as a couple
$(\al_{s},\th_{s})$.
For a given sector $(\al_{s},\th_{s})$, we define
the support $\Si(\al_{s},\th_{s})$  to be the support of the
function $\chi^{\th_s}_{\al_{s}}(\th)$. Inside this support
$|\th-\th_s|\leq (1/2) \al_s^{-{1/4}}$.

Now, in order to exploit momentum conservation at each vertex and subgraph,
we need to  decompose each half-line  (either loop, tree or external)
a certain number of times into sectors
with different values of $\al_s$,  starting from larger sizes
(hence smaller $\al_s$) and then refining them into smaller ones.

This process requires to define a sequence of scales for each
line. These scales roughly speaking represent all scales $i$ for which
the half line is external to the subgraph $g_i$ and
$|eg_i({\cal C})|\leq 10$
 (as we can exploit momentum conservation only in this case),
plus a last scale, characteristic of the line and the class ${\cal C}$.
The subgraph $g_r$ requires a particular treatment: its external lines are
the only real external lines of the whole graph, hence
we can always exploit momentum conservation, even if $2p=|eg_r|>10$.

\subsection{Choice of scales $\al_s$ for each half-line}

Let us introduce an index $h$ which parametrizes
 loop, tree and external half-lines. The sum over sector
choices will be  done inductively, from  the root
towards the leaves.
We then choose as root vertex the external vertex $x_{e_1}$, and as root
the test function   $\phi_{e_1}$.
Now we denote the two  half-lines belonging to
the tree line $l_i$ as $h^L_{i}$ ($h$ left) and $h^R_{i}$
($h$ right) in such a way
that $h^R_{i}\to h^L_{i}$ is oriented towards the root vertex.
Hence we define
$\tree_L$ and $\tree_R$ as the set of tree half-lines of
left and right type  respectively.

Remark that, for any subgraph $g_k$ with $k\neq r$, (as $eg_r=2p$ then
there is no tree external line) there is at most
one tree  half-line $h_i\in et_k\cap \tree_R$
(that we call $h_{k}^{root}$)  going towards
the root. If $e_1\in ee_i$ all tree external half-lines belong to $\tree_L$
and we put  $h_{k}^{root}= e_1$. The sector of this half-line is kept fixed
in the sum over sector choices until scale $0$.
In the same way the sector of each tree right half-line $h_i^R$ is
kept fixed in the sum
over sector choices until scale $i$;
by momentum conservation along the tree line
$l_i$ this sector is then equal to that of $h^L_i$.
Therefore for each tree line $l_i$ we perform sector decoupling and
sector sums only  for $h^L_{i}$ (as $h^R_{i}$ is automatically fixed by
$h^L_{i}$).
In the following, $h^R$ will appear only as $h_i^{\rm root}$ for some subgraph
$g_i$, hence to simplify notation we write simply $h_i$ for $h_i^L$.

Given the class ${\cal C}$ we define a natural scale
$i(h)$ associated to each $h\in L\cup \tree_L\cup E$
\begin{itemize}
\item{} For the  left half-line   belonging to  the tree line $l_{i}$
obviously $ i(h_{i}) = i$.
\item{}
For a loop  half-line $h=a$ we choose $i(h) = M(a, {\cal C})$
(this choice avoids the ``logarithmic divergence'' associated to momentum
conservation in 2 dimensions, see [FMRT1], lemma 2).
\item{} For all external lines we choose $i(e)=0$ which is the
band to which they belong. This means that we cut them in sectors
of size $\al_0^{-{1\over 4}} := \La_T^{1\over 2}$.
\end{itemize}
We introduce then a growing sequence of indices $j_{h,1}=i(h),...,
j_{h,n_h}=i_{v_h}$ such that each scale $j_{h,r}$
of the sequence corresponds to
a refining of that half-line in sectors  of size
$1/\al_{j_{h,r}}^{1/4}=\La^{1/2}(w_{j_{h,r}})$.
Remark that the lowest refining scale is $i(h)$.

The choice of these indices is the following:
a half-line $h\in \tree_L\cup L\cup E$
is refined at scale $j=i(h)$ and at all scales $j$ such that
 $h\in eg_i({\cal C})$ for some level $i$
with $j={\cal A}(i)$ and such that $ |eg_i({\cal C})|\leq 10$.

This multiple decomposition has to be adapted to the different bounds satisfied
by tree, loop, and external lines.

\subsubsection{Tree lines}
As explained above, we introduce the multi-sector decomposition only for
the left half-line of $l_i$, $h_i$.
We must ensure that the spatial decay of the tree line $l_i$
depends only on the finest sector (at level $i$),
hence,  we apply the identity (\ref{sec1}) just one time, at the scale  $i$.

We then decompose each tree left half-line on larger sectors introducing the
identity
\be
1 = \left [{4\over 3}\al_{j_{h_i,r}}^{1/4} \right ]
\int_0^{2\pi}
d\th_{h_i,r} \; \chi^{\th_{h_i,1}}_{\al_{j_{h_i,r}}}(\th_{h_i,r})\ .
\label{dec}\ee
This actually selects $\th_{h_i,r}$ to be in
a sector of size $\La^{1/2}(w_{j_{h_i,r}})$ around  $\th_{h_i,1}$.
Hence, for the half-tree line $h_i\in\tree_L$
the complete decomposition is:
\bqa
1&=&
  \left [{\scriptstyle{4 \over 3}\al_{j_{h_i,1}}^{1/4} }\right ]
\int_0^{2\pi}  d\th_{h_i,1} \; \chi^{\th_{i}}_{\al_{j_{h_i,1}}}
(\th_{h_i,1}) \left \{ \prod_{r=2}^{n_{h_i}} \left [{\scriptstyle{4 \over 3}
\al_{j_{h_i,r}}^{1/4}}\right ]\int_0^{2\pi}
d\th_{h_i,r} \; \chi^{\th_{h_i,1}}_{\al_{j_{h_i,r}}}(\th_{h_i,r})\right \}
\no\\&=&
\left [{\scriptstyle{4 \over 3}\al_{j_{h_i,n_{h_i}}}^{1/4}}\right ]
\int_0^{2\pi}  d\th_{h_i,n_{h_i}}
\left [{\scriptstyle {4 \over 3}\al_{j_{h_i,n_{h_i}-1}}^{1/4}}\right]
\int_{\Si_{j_{h_i,n_{h_i}}}}
d\th_{h_i,n_{h_i}-1}\;\; ...\no\\
&&
\left [{\scriptstyle {4 \over 3}\al_{j_{h_i,2}}^{1/4}}\right ]
\int_{\Si_{j_{h_i,3}}}
d\th_{h_i,2}
\left [[{\scriptstyle{4 \over 3}\al_{j_{h_i,1}}^{1/4}}\right ]
\int_{\Si_{j_{h_i,2}}}
d\th_{h_i,1}
\left [ \prod_{r=2}^{n_h}
\chi^{\th_{h_i,1}}_{\al_{j_{h_i,r}}}(\th_{h_i,r})\right ]
\chi^{\th_{i}}_{\al_{j_{h_i,1}}}(\th_{h_i,1})
\no\\
\label{dec1}\eqa
where we defined sectors twice as large as the previous ones:
\be
\Si_{j_{h,r}} :=
 \Si(\al_{j_{h,r}}/2^4,\th_{h,r}) \equiv
\{\th  \; |\; |\th_{h,r}-\th|\leq \La^{1/2}(w_{j_{h,r}})\} \ .
\label{sec3}\ee
Indeed the integration domain for $\th_{h_i,r}$, $r\ge 2$,
can be restricted to
$\Si_{j_{h_i,r+1}}$ if we observe that the product
$\chi^{\th_{h_{i},1}}_{\al_{j_{h_{i},r}}}(\th_{h_{i},r})
\chi^{\th_{h_{i},1}}_{\al_{j_{h_{i},r+1}}}(\th_{h_{i},r+1})$
can be non zero only if $\th_{h_{i},r}  \in \Si_{j_{h_i,r+1}}$.
This is also true for $r=1$ since the single function
$\chi^{\th_{h_{i},1}}_{\al_{j_{h_{i},2}}}(\th_{h_{i},2})$ is non zero
only if $|\th_{h_{i},1}-\th_{h_{i},2}| \le {1\over 2}\La^{1/2}(w_{j_{h,2}})$,
which implies $\th_{h_{i},1} \in \Si_{j_{h_i,2}}$.

Finally we remark that for each $r \ge 1$, $\th_{i}\in \Si_{j_{h_{i},r}}$,
where $\th_{i}$ is the angular variable for the momentum of the propagator
of line $i$.

\subsubsection{External and loop half-lines}
External test functions enter in spatial integration too, hence we perform
the sector decomposition in the same way as for tree left half-lines.
Loop lines are not used in spatial decay, and there is no sector conservation
along the line, as we do not know exactly which loop fields are contracted.
Hence we can decompose them as we want. To simplify notation, we treat them
exactly in the same way as the tree left half-lines.

Hence the expression (\ref{convpo}) for the  convergent
part of the vertex function becomes:
\bqa
&&\hspace{-0.7cm}\sum_{n=1}^\infty \frac{\la^n}{n!} \sum_{CTS}
\sum_{u-\tree}\sum_{\cal L}\sum_{E}\sum_{\Om}\sum_{{\cal C}_{\rm c}}
\vep(\tree, \Om) \int_{w_T\le w_{{\cal A}(i)} \le  w_{i}\le 1}
\prod_{q=1}^{n-1}     dw_q   \label{conv1}\\
&&\prod_{h\in L\cup\tree_L\cup E}
\left\{ [{\scriptstyle {4 \over 3} \La^{-{1\over 2}}(w_{j_{h,n_h}})}]
\int_0^{2\pi}  d\th_{h,n_h}
[ {\scriptstyle{4 \over 3}\La^{-{1\over 2}}(w_{j_{h,n_h-1}})}]
\int_{\Si_{j_{h,n_h}}}
d\th_{h,n_h-1} \right .\no\\
&&...\;\;[{\scriptstyle {4 \over 3}\La^{-{1\over 2}}(w_{j_{h,1}})}]
\left .\int_{\Si_{j_{h,2}}}
d\th_{h,1}
\left [\prod_{r=2}^{n_{h}}
\chi^{\th_{h,r}}_{\al_{j_{h,r}}}(\th_{h,1})\right ]
 \right \}\no\\
&&\int
d^3x_1...d^3x_{n}\;\;
\phi^{\La_T}_1(x_{i_{1}},\th_{e_1,1})\;...\;
\phi^{\La_T}_{2p}(x_{j_{p}},\th_{e_{2p},1})\no\\
&&\biggl[\prod_{q=1}^{n-1}  C^{w_q}
( x_{q}, {\bar x}_{q}, \th_{h,1}) \biggr]
\det{\cal M}'({\cal C},E,\{\th_{a,1}\})\ ,
\no\eqa
where for $1\le l\le 2p$
\be
\phi^{\La_T}_l(x,\th_{e_{l},1}):= \frac{1}{(2\pi)^2 }
\int d^3k \; e^{ikx}\;\chi^{\th_{e,1}}_{\al_0}(\th)\
\hat{\phi}(k) \left. [u(r/\La^2_T)]\right |_{r=k_0^2+e^2(\vec{k})},
\label{phisec}\ee
\be
C^{w_q}(\bar{x}_{q}, x_{q}, \th_{h,1})
:=  \frac{1}{(2\pi)^2 }\int d^3k \; e^{ik(x_{q}-\bar{x}_{q})}
C^{w_q}(k)
\chi^{\th_{h,1}}_{\al_{j_{h,1}}}(\th)\ ,
\label{treesec}\ee
and the coefficients of the matrix ${\cal M}'({\cal C},E,\{\th_{a,1}\} )$ are
\bqa
 &&\hspace{-0.6cm}{\cal M}'({\cal C},E,\{\th_{a,1}\} )_{x_f,x_g} :=
\label{loopsec}\\
&&\hspace{-0.7cm} \frac{1}{(2\pi)^2 }\int d^3k\; e^{ik(x_f-x_g)} C(k)
\sum_{q=1}^n \eta_{a(f)}^q\eta_{a(g)}^q u^q(k)
W^q_{v_f,v_g}
\left [
\chi^{\th_{a(f),1}}_{\al_{j_{a(f),1}}}(\th)\right ]
\left [
\chi^{\th_{a(g),1}}_{\al_{j_{a(g),1}}}(\th)\right ]\, .\no
\eqa
Remark that the sums over sectors have been taken out of the
determinant by multi-linearity, and that we used $\chi^{\th_{1}}_{\al_1}
(\th)=\chi^{\th}_{\al_1}
(\th_{1}) $.

Now we want to exploit momentum conservation.
At each subgraph $g_i$ with $i=r$ or $|eg_i({\cal C})|\leq 10$
we refine all external
lines in sectors  at the scale ${\cal A}(i)$, except for the
half-line $h_i^{root}$ which is fixed in a sector of size
$\La^{{1\over 2}}(w_j)\leq \La^{{1\over 2}}(w_{{\cal A}(i)})$
(for some $0 \leq j\leq {\cal A}(i)$).
Actually the volume of integration for the new
sectors is restricted by momentum conservation.
To take into account these effects we insert in the expression above
\be
1 =\Upsilon\left (\th_{h_i^{root}},
\{\th_{h,r(i)}\}_{h\in eg^\ast_i} \right ) +
\left [1- \Upsilon\lp\th_{h_i^{root}},
\{\th_{h,r(i)}\}_{h\in eg^\ast_i}\rp  \right  ]\ , \label{momcons}
\ee
where we defined $r(i)$ as the number of refinements we have done on
the half-line $h$ until ${\cal A}(i)$ (this means
$j_{h,r(i)}={\cal A}(i)$). We also
set $eg_i^\ast:= eg_i\backslash \{h_i^{root}\}$  and define the function
$\Upsilon$ to be 0 if the set of selected sectors  is forbidden by
momentum conservation at this subgraph, and we define $\Upsilon$ to be 1
otherwise. Therefore after insertion of (\ref{momcons})
the term $1-\Upsilon$, forbidden by momentum conservation,
gives a zero contribution.
Hence we can insert freely in (\ref{conv1}) the product
\be
\prod_{g_i|\ i=r \; {\rm or} \; |eg_i({\cal C})|\leq 10}
 \Upsilon\lp\th_{h_i^{root}},
\{\th_{h,r(i)}\}_{h\in eg^\ast_i}\rp.
\ee
In this way we exploit momentum conservation at each subgraph, but we
still have to exploit it at each vertex.
For that we need some additional notation. We call $H(v)$ the set of
half-lines hooked to $v$ (and  $|H(v)|$  its cardinal).
We define $H^\ast(v):= H(v)\backslash h_v^{\rm root}$ where   $h_v^{\rm root}$
is the  half-line  going towards the root.
Remark that the scale $i_v$ is the largest scale of refinement
for each of the elements of $H^\ast(v)$:
$i_v=j_{h,n_h}$, $\forall h\in H^\ast(v)$.
Again  we  can insert the function
 $\Upsilon\left (\th_{h_v^{root}},
\{\th_{h,n_h}\}_{h\in H^\ast(v)} \right )$, which is zero when
the sectors are not permitted by
momentum conservation at vertex $v$. Hence, by the same argument as above,
we can freely insert in (\ref{conv1})
\be
\prod_{v}
 \Upsilon\lp\th_{h_v^{root}},
\{\th_{h,n_h}\}_{ h\in H^\ast(v)}\rp\ .
\ee

\section{Main result and Bounds}
\resetequ 
Now we have all the elements to perform the bounds. We insert absolute
values inside the sums and integrals and obtain the inequality
\bqa
\lefteqn{
|\Ga_{2p\; {\rm conv.}}^{\La\Lazero}|\leq \sum_{n=1}^\infty \frac{|\la|^n}{n!}
\sum_{CTS}\sum_{u-\tree}\sum_{\cal L}\sum_{E\; \Om}
\sum_{{\cal C}_{\rm c}} \int_{w_T\le w_{{\cal A}(i)} \le  w_{i}\le 1}    
\prod_{q=1}^{n-1} dw_q }   \label{conv2}\\
&&
\prod_{h\in L\cup\tree_L\cup E}
 \left \{ [{\scriptstyle {4 \over 3}\La^{-{1\over 2}}(w_{j_{h,n_h}})}]
\int_0^{2\pi}  d\th_{h,n_h}\;\;
 [{\scriptstyle{4 \over 3}\La^{-{1\over 2}}(w_{j_{h,n_h-1}})}]
\int_{\Si_{j_{h,n_h}}}
d\th_{h,n_h-1} \right .\no\\
&&...\;\;[{\scriptstyle {4 \over 3}\La^{-{1\over 2}}(w_{j_{h,1}})}]
\left .\int_{\Si_{j_{h,2}}}
d\th_{h,1}
\left [\prod_{r=2}^{n_{h}}
 \chi^{\th_{h,r}}_{\al_{j_{h,r}}}(\th_{h,1})
\right ] \right \}
\no\\
&&
\prod_{g_i|\ i=r\; {\rm or}\atop |eg_i({\cal C})|\leq 10}
\Upsilon\lp\th_i^{root},\{\th_{h,r(i)}\}_{h\in eg^\ast_i}
 \rp \;
\prod_{v}  \Upsilon\lp\th_{h_v^{root}},
\{\th_{h,n_h}\}_{ h\in H^\ast(v)} \rp\no\\
&&\int
d^3x_1...d^3x_{n}\;\;
|\phi^{\La_T}_1(x_{i_{1}},\th_{e_1,1})|...
|\phi^{\La_T}_{2p}(x_{j_{p}},\th_{e_{2p},1})|
\no \\
&&
\biggl[\prod_{q=1}^{n-1} | C^{w_q}
( x_{q}, {\bar x}_{q}, \th_{h,1})| \biggr]
|\det{\cal M}'({\cal C},E,\{\th_{a,1}\})|
\no\eqa

Actually we prove the following theorem (more precise than Theorem 1):
\begin{theor}
Let $\vep>0$, $\La_0=1$ and $T\ge 0$ be fixed. The series (\ref{conv2}) is 
absolutely
convergent for $|\la| \leq c$, $c$ small enough. This convergence is uniform 
in $\La$, then the IR limit 
$\Ga_{2p,conv}^{\La_0}=\lim_{\La\rightarrow 0}\Ga_{2p,conv}^{\La\La_0}$ 
exists and satisfies the bound:
\bqa
&&\hspace{-0.6cm}|\Ga^{\La_0}_{2p>4,conv}(\phi_1^{\La_T},..., 
\phi_{2p}^{\La_T})| \leq \\
&&\qquad
 K_0\; {\scriptstyle ||\phi_1||_{_1}
\prod_{i=2}^{2p} ||\hat{\phi}_i||_\infty }
{T^{{7\over 4}2p -{1\over 2}}\over 2p-4} \ [K_1(\vep)]^p\  \lp p!\rp^2 \;
K(c)
\; e^{-(1-\vep)\La_T^{1\over s}d^{1\over s}_{\tree}(\Om_1,...\Om_{2p})}\no
\eqa
\[
|\Ga_{4,conv}^{\Lazero}(\phi_1^{\La_T},..,\phi_4^{\La_T})|\leq
 K'_0 \,
{\scriptstyle ||\phi_1||_{_1}
\prod_{i=2}^{4} ||\hat{\phi}_i||_\infty }
T^{13\over 2}  |\log T|  
K(c)
\; e^{-(1-\vep)\La_T^{1\over s}d^{1\over s}_{\tree}(\Om_1,...\Om_{4})}
\]
\[\hspace{-1.2cm}
|\Ga_{2,conv}^{\Lazero}(\phi_1^{\La_T},\phi_2^{\La_T})|\leq
  K''_0\;
{\scriptstyle ||\phi_1||_{_1}
 ||\hat{\phi}_2||_\infty }
T^2 \ 
K(c)
\; e^{-(1-\vep)\La_T^{1\over s}d^{1\over s}_{\tree}(\Om_1,\Om_{2})}
\]
where $\Om_i$ is the compact support of $\phi_i$, $K_1(\vep)$  is a constant 
dependent 
from $\vep$, $K(c)$ is a function of $c$  that tends to zero when $c$ tends to 
zero, and $s$ is the Gevrey index of our cutoff function $u$ (we assume
that $1<s<2$). Finally we defined
\bqa
d_{\tree}(\Om_1,...\Om_{2p}) &:= &\inf_{x_i\in\Om_i} 
d_{\tree}(x_1,...,x_{2p})\ ,\no\\
d_{\tree}(x_1,...,x_{2p})   &:= &\inf_{u-\tree} 
\sum_{l\in\tree} |{\bar x}_l-x_l|\ ,
\eqa
where in the definition of $d_{\tree}(x_1,...,x_{2p})$
(called the tree distance of $x_1,...x_{2p}$) 
the infimum over $u-\tree$ is taken over all unordered trees 
(with any number of vertices) connecting $x_1,...x_{2p}$.
\end{theor}

\subsection{Loop determinant}

To bound the loop determinant we apply Gram's inequality, which states that if 
$M$ is a $n\times n$ matrix whose elements $M_{ij}= <f_i,g_j>$
are scalar products of vectors $f_i$, $g_j$  in a Hilbert  space, then
$|\det M|\leq \prod_{i=1}^n ||f_i||\;  \prod_{j=1}^n ||g_j||$.

\begin{lemma}
The matrix ${\cal M}'({\cal C})$ satisfies the following Gram inequality:
\bqa
&&|\det {\cal M'}({\cal C})| \leq
 \prod_f ||F_f||_{\cal C}\;  \prod_g ||G_g||_{\cal C}\label{gram}\\
&&\quad = \prod_f \left [
\frac{1}{(2\pi)^2}\int d^3k\; u_{\cal C}^f(k)  |F_f(k)|^2
\right ]^{\frac{1}{2}}
 \prod_g \left [
\frac{1}{(2\pi)^2}\int d^3k\; u_{\cal C}^g(k)  |G_g(k)|^2
\right ]^{\frac{1}{2}}
\no\eqa
where the cut-off  $u_{\cal C}^a(k)$  is defined by:
\be
u_{\cal C}^a (k) :=  \left [  u\lp
\frac{k_0^2+e^2(\vec{k})}
{\La^2({w}_{M(a,{\cal C})})}\rp -
 u\lp \frac{k_0^2+e^2(\vec{k})}
{\La^2({w}_{{\cal A}(m(a,{\cal C}))})}\rp \right ].
\ee
\end{lemma}
\paragraph{Proof}
The proof is identical to that of 
Lemma 4 in [DR1]. The only difference is that here we have partial order 
instead of the total order in [DR1]. We just resume it for
completeness.  
We observe that the matrix element  (\ref{loopsec}) can be written
as
\be
\frac{1}{(2\pi)^2}
\int d^3k \; F_f(k)\; G_g^*(k)\;
\sum_{q=1}^n \; W^q_{v_f,v_g} \; u^q(k)\; \eta^q_{a(f)}\;
\eta^q_{a(g)}
\ee
where we defined
\be
F_f(k)= e^{ix_f k}
\chi^{\th_{f,1}}_{\al_{j_{f,1}}}(\th)
\frac{1}{(k_0^2+e^2(\vec{k}))^{\frac{1}{4}}}\quad
G_g(k)= e^{ix_g k}
\chi^{\th_{g,1}}_{\al_{j_{g,1}}}(\th)
\frac{(ik_0+e(\vec{k}))}{(k_0^2+e^2(\vec{k}))^{\frac{3}{4}}}\ .
\ee
We introduce the matrix 
\be
 {\cal W}^{q}_{v,a;v',b}\; := \; R^q_{a,b}  \; W^q_{v,v'} := 
\eta^q_a \; \eta_b^q \; W^q_{v,v'}
\ee
for $v$, $v'$ belonging to the set of $n$ vertices, $a$, $b$ to the set of
$2n+2-2p$ loop half-lines (fields and anti-fields). 
Both  $R^q_{a,b}$  and $W^q_{v,v'}$ can be written (modulo permutation of 
field and vertex indices) as block diagonal positive matrices or sums
of matrices of the type
\be
\lp \ba{cc}
1_k & 0 \\
0 & 0 \\
\ea\rp
\ee
where $1_k$ is a $k\times k$ matrix with all elements equal to 1. 
Then ${\cal W}^q$ is positive, $\sum_q  u^q\; {\cal W}^q$ is positive too
and there exists a positive  matrix $U$ defined by
\be
\sum_{w,c}\; U_{v,a;w,c}\; U_{w,c;v',b}\;
:=\;\sum_q\; u^q\; {\cal W}^{q}_{v,a;v',b} \ .
\ee
The determinant  can be written as the scalar
product of two functions
\be
 {\cal M}'_{fg}= \frac{1}{(2\pi)^2}
 \int d^3k \;\sum_{v', s}  {\cal F}^{f}_{v's}\;
\; {\cal G}^{g*}_{v's} \;
= <\vec{{\cal F}}^f,\vec{{\cal G}}^g> \ ,
\ee
where we defined
\be
{\cal F}^f_{v's}(k) = F_f(k)\;  U_{v',s;v(f),a(f)}\quad , \quad 
{\cal G}^g_{v's}(k) = G_g(k) \; U_{v',s;v(g),a(g)} \ .
\ee
Applying  Gram inequality we obtain (\ref{gram}). 
\qed

With these definitions, the norms of $F_f$ and $G_g$
satisfy the bounds
\bqa
||F_f||_{\cal C}& \leq &
  K\;\La^{\frac{1}{4}}(w_{M(f,{\cal C})})\;    [\La({w}_{M(f,{\cal C})})
-\La({w}_{{\cal A}(m(f,{\cal C}))})]^{\frac{1}{2}}\no\\
||G_g||_{\cal C}& \leq &
  K\;\La^{\frac{1}{4}}(w_{M(f,{\cal C})})  [\La({w}_{M(g,{\cal C})})
-\La({w}_{{\cal A}(m(g,{\cal C}))})]^{\frac{1}{2}}\ .   \label{loopb1}
\eqa
Indeed let us bound for instance the norm of $F_f$:
\bqa
&&\hskip-1.1cm||F_f||_{\cal C}^2= \frac{1}{(2\pi)^2}\int d^3k\;
\frac{[
\chi_{\al_{j_{f,1}}}^{\th_{f,1}}(\th)]^2}
{[k_0^2+e^2(\vec{k})]^{1 \over 2}}
\left [u\lp \frac{k_0^2+e^2(\vec{k})}{\La^2({w}_{M(f,{\cal C})})} \rp
-u\lp\frac{k_0^2+e^2(\vec{k})}{\La^2({w}_{{\cal A}(m(f,{\cal C}))})}
\rp \right ]\no\\
 &= &
\int_{\Lainv({w}_{M(f,{\cal C})})}^{\Lainv
({w}_{{\cal A}(m(f,{\cal C}))})} d\al
 \frac{1}{(2\pi)^2}\int  d^3k\; 
[\chi_{\al_{j_{f,1}}}^{\th_{f,1}}(\th)]^2\;
\left .\left [
- x^{1 \over 2} \; u'[\al x]\right ]\right |_{x= k_0^2+e^2(\vec{k})} \no \\
&\leq  &
\int_{\Lainv({w}_{M(f,{\cal C})})}^{\Lainv({w}_{{\cal A}(m(f,{\cal C}))})}
d\al \;\frac{1}{\beta}|{\rm S}|\;
\sup_{\rm S}\left [
  \chi_{\al_{j_{f,1}}}^{\th_{f,1}}(\th)   \left .\left (
 - x^{1 \over 2}
u'[\al x ]\right )\right |_{x= k_0^2+e^2(\vec{k})} \right ] \no\\
&\leq &K \La^{\frac{1}{2}}(w_{M(f,{\cal C})}) \;
\int_{\Lainv({w}_{M(f,{\cal C})})}^{\Lainv({w}_{{\cal A}(m(f,{\cal C}))})}
d\al \;
\al^{-\frac{3}{2}}  \no\\
&\leq &
K\;\La^{\frac{1}{2}}(w_{M(f,{\cal C})})\;
[\La({w}_{M(f,{\cal C})})-\La({w}_{{\cal A}(m(f,{\cal C}))})]\ , 
\label{loopb2}
\eqa
where $K$ is some constant, $S$ is the set in momentum space selected
by the cut-offs $\chi$ and $u'$, and we applied the bounds:
\bqa
&&
[\chi_{\al_{j_{f,1}}}^{\th_{f,1}}(\th)]^2
\leq  \chi_{\al_{j_{f,1}}}^{\th_{f,1}}(\th) \no\\
&&\sup_{\rm S}\left [
   \chi_{\al_{j_{f,1}}}^{\th_{f,1}}(\th)  \left .\left (
 - x^{1 \over 2}
u'[\al x ]\right )\right |_{x= k_0^2+e^2(\vec{k})} \right ]
\leq \;K \al^{-\frac{1}{2}}\no\\
&&|{\rm S}| \leq\; \beta \La^{\frac{1}{2}}(w_{M(f,{\cal C})})\; \al^{-1} \  .
\eqa
Finally the loop determinant is bounded by
\be
 |\det{\cal M}'({\cal C},E,\{\th_{a,1}\})|\leq  K^{n} \prod_{a\in L}
\;\La^{\frac{1}{4}}(w_{M(a,{\cal C})})\    [\La({w}_{M(a,{\cal C})})
-\La({w}_{{\cal A}(m(a,{\cal C}))})]^{\frac{1}{2}} \ .
\ee
This bound no longer depends from $\{\th_{a,r}\}$ or $E$.

\subsection{Spatial integrals}

To perform spatial integration we use the decay of tree lines.
The test functions are taken out of the integral and bounded by their
$L_\infty$ norm, except $\phi^{\La_T}_1$ which is used to perform the 
integration over the root $x_1$.

\bqa
&&\int d^3x_1...d^3x_n
|\phi^{\La_T}_1(x_{i_{1}},\th_{e_1,1})|...
|\phi^{\La_T}_{2p}(x_{j_{p}},\th_{e_{2p},1})|
 \prod_{q=1}^{n-1}
| C^{w_q}
( x_{q}, {\bar x}_{q}, \th_{h_q,1})|
\no\\
&& \leq {\scriptstyle ||\phi^{\La_T}_1(\th_{e_1,1})||_{_{\scriptstyle 1}}
\prod_{i=2}^{2p} ||\phi^{\La_T}_i(\th_{e_i,1})||_\infty}
 \int d^3x_2...d^3x_n
 \prod_{q=1}^{n-1}
| C^{w_q}( x_{q}, {\bar x}_{q}, \th_{h_q,1})|\ . \no\\
\label{spint0}\eqa

We now bound the norms of the test functions and the spatial decay of the
tree propagators.

\subsubsection{Test functions}

Each test function $\phi^{\La_T}_i$ ($i=1,...2p$) obeys the bound
\bqa
\lefteqn{||\phi^{\La_T}_i(\th_{e_i,1})||_\infty
= \sup_{x} \left |
\frac{1}{(2\pi)^2}\int d^3k\;
e^{ikx}
\left [  \chi_{\th_{e_i,1}}(\th)\right ] \hat{\phi}_i(k)
\left. [u(r/\La^2_T)]\right |_{r=k_0^2+e^2(|k|)}
\right |}\no\\
&&\leq
\frac{1}{(2\pi)^2}\int d^3k\;
\chi_{\th_{e_i,1}}(\th)
\;\;\left |\hat{\phi}_i(k)\right |\left. [u(r/\La^2_T)]
\right |_{r=k_0^2+e^2(|k|)}
\no\\
&&
\leq
||\hat{\phi_{i}}||_\infty
\frac{1}{(2\pi)^2}\int d^3k\;
\chi_{\th_{e_i,1}}(\th)
\left. [u(r/\La^2_T)]\right |_{r=k_0^2+e^2(|k|)}
\leq K\;\La^{5\over 2}_{T}\;  ||\hat{\phi}_i||_\infty\ , \no\\
 \eqa
where in the third line we used
$1/\al_{j_{e,1}}=\La^2(w_{j_{e,1}})=\La^2_T$ $\forall e$.
For the test function hooked to the root, we need to perform a different
bound. We write (recalling our convention (\ref{convention}) of integration, 
which includes that the imaginary time variable is integrated on a circle):
\be
||\phi^{\Lambda_T}_1(\th_{e_1,1})||_{_{\scriptstyle 1}}
\;=\;\int d^3x d^{3} y\; \left |
\phi_1(x,\th_{e_1,1})\eta_{\th_{e_1,1}}(x-y) \right | \label{box1}
\ee 
where we defined $\eta_{\th_{e_1,1}}(x)$ as the Fourier transform of
\be
\hat{\eta}_{\th_{e_1,1}}(k) =
\chi_{\th_{e_1,1}}(\th) \left. [u(r/\La^2_T)]\right |_{r=k_0^2+e^2(|k|)}
\ee
\begin{lemma}
$\eta_{\th_{e_1,1}}(z) $ decays as:
\be
|\eta_{\th_{e_1,1}}(z) |\leq K\;\sum_m {\La^{5/2}_T\over
[1+\La^2_T |z_0+2m\beta|^2+\La^2_T|z_r|^2+\La_T|z_t|^2]^2}
\label{dec2}\ee
where $z_r$ and $z_t$ are the radial and tangential components of
$\vec{z}$ relative to  the sector center $\th_{e_1,1}$.
\end{lemma}
\paragraph{Proof}
This is a standard duality between direct
and momentum space. However since the time variable is periodic
we cannot immediately derive with respect to the 0-th component of the 
momentum. Instead we can derive first the decay of the $T=0$ analog
$\eta^{0}_{\th_{e_1,1}}(z)$ of the function $\eta_{\th_{e_1,1}}(z) $.
We write
\bqa
F &=& [1+\La^2_T |z_0|^2+\La^2_T|z_r|^2+\La_T|z_t|^2]^2 \;
|\eta^{0}_{\th_{e_1,1}}(z) | \no \\
&=& \frac{1}{(2\pi)^2}
\left |\int d^3k\; [1+\La^2_T |z_0|^2+\La^2_T|z_r|^2+\La_T|z_t|^2]^2\;
e^{ikz}  \;
  \; \eta_{\th_{e_1,1}}(k)\right |\no\\
&& \leq K\int d^3k  \;
\left |[1-\La^2_T \partial^2_{k_0}-\La^2_T\partial^2_{k_r}-
\La_T\partial^2_{k_t}]^2\;
 \;  \eta_{\th_{e_1,1}}(k)\right |\no\\
&&\leq K \; \La^{5\over 2}_T \ .
\eqa
Therefore
\[
|\eta^0_{\th_{e_1,1}}(z) |\leq K\;{\La^{5\over 2}_T\over
[1+\La^2_T |z_0|^2+\La^2_T|z_r|^2+\La_T|z_t|^2]^2}\ .
\]
Now applying (\ref{copie1}) we can end the proof.
\qed

We introduce the spatial decay (\ref{dec2}) in (\ref{box1}) to obtain:
\bqa
&&\hspace{-.5cm} ||\phi_1^{\Lambda_T}(\th_{e_1,1})||_{_{\scriptstyle 1}} \leq
  ||\phi^0_1(y)||_{_{\scriptstyle 1}} \\
&&\hspace{-.5cm}\sum_m
\int_{-{1\over T}}^{1\over T} dz_0\int d^2z\;
\; { K\La^{5\over 2}_T\over
(1+\La^2_T |z_0+{2m\over T}|^2+\La^2_T|z_r|^2+\La_T|z_t|^2)^2}\no\\
 &&\hspace{-.5cm}= ||\phi^0_1(y)||_{_{\scriptstyle 1}}
\int dz_0\int d^2z\;
\; { K\La^{5\over 2}_T\over
(1+\La^2_T |z_0|^2+\La^2_T|z_r|^2+\La_T|z_t|^2)^2}
\le K ||\phi_1(y)||_{_{\scriptstyle 1}}  .\no
\eqa
where we performed the change of variable
$z_0+{m\over T}\rightarrow z_0$.

\subsubsection{Spatial decay of tree lines}

We consider now tree line propagators and prove that they  decay as
Gevrey functions of class $s$ where
$s$ is the Gevrey index of our initial cutoff $u$.
\bqa
\lefteqn{| C^{w_q}
( \de x_q,0, \th_{h_q,1})|
\leq}\label{spdec0}\\
&& K \; \frac{\La_0^2-\La^2}{\La^4(w_q)}
\;\La^{{1\over 2}}(w_q) \;\La^3(w_q)\; e^{-a\left [
|(\de x_q)_0 \La(w_q)|^{{1\over s}}+|(\de x_q)_r \La(w_q)|^{{1\over s}}+
 |(\de x_q)_t \La^{{1\over 2}}(w_q)|^{{1\over s}}
\right ]}
\no\eqa
where we applied translational invariance, $\de x_q:=x_q-{\bar x}_q$, 
$(\de x_q)_r$ and $(\de x_q)_t$ are the radial and tangential components of
$\vec{x}$ relative to  the sector center $\th_{h,1}$, 
$K$ and $a$ are some positive constants.
Remark that the smallest sector governs the spatial decay rate.

To prove this formula we study, as for the test function $\phi_1$,
the propagator at $T=0$ $C_0^{w_q}$. Using the properties of Gevrey 
functions with compact support, $C_c^{w_q}$ satisfies (\ref{spdec0}) too 
(see Appendix A). Then applying (\ref{copie1}) achieves the proof of
(\ref{spdec0}).

\subsubsection{Bound} Now we can complete the bound on (\ref{spint0}). 
But before that, in order to extract the exponential decay between the 
test functions supports of Theorem 3, we take out a fraction  $(1-\vep)$ 
of  the exponential decay of each tree line in (\ref{spdec0}). This factor 
is bounded by 
\be
\prod_{q=1}^{n-1}  e^{-a (1-\vep)\left (
|(\de x_q)_0 \La(w_q)|^{1\over s}+|(\de x_q)_r \La(w_q)|^{1\over s}+
 |(\de x_q)_t \La^{1\over 2}(w_q)|^{1\over s}\right )}
\leq e^{-a\;(1-\vep)\;\La_T^{1\over s}\;  d^{1\over s}_T(\Om_1,...,\Om_{2p}) }.
\label{tests}
\ee
We keep the remaining fraction  $\vep$ of the decay to perform spatial 
integration:
\bqa
&&\int d^3x_2...d^3x_n
 \prod_{q=1}^{n-1}
 e^{-a \vep\left (
|(\de x_q)_0 \La(w_q)|^{1\over s}+|(\de x_q)_r \La(w_q)|^{1\over s}+
 |(\de x_q)_t \La^{1\over 2}(w_q)|^{1\over s}\right )} \no\\
&\leq&
\prod_{q=1}^{n-1} \left [
\int d^3x\;  e^{-a \vep \left (
|x_0 \La(w_q)|^{1\over s}+|x_r \La(w_q)|^{1\over s}+
 |x_t \La^{1\over 2}(w_q)|^{1\over s}\right )}\right ]
\no\\
& \leq & \prod_{q=1}^{n-1} \frac{1}{\La^{5\over 2}(w_q)} 
\int d^3u \; e^{-a \vep \left [
u_0^{1\over s}+u_1^{1\over s}+u_2^{1\over s}
\right ]} \;    
\leq  \; K  \prod_{q=1}^{n-1} \frac{1}{\La^{5\over 2}(w_q)}
\eqa
and eq(\ref{spint0}) is bounded by
\be
 K {\scriptstyle ||\phi_1||_{_1}}
 {\scriptstyle \left [\prod_{i=2}^{2p}   ||\hat{\phi}_i||_\infty \right ]
\lp\La^{5\over 2}_T\rp^{(2p-1)}}
e^{-a\;(1-\vep)\;\La_T^{1\over s}\;  d^{1\over s}_T(\Om_1,...,\Om_{2p}) }.
 \prod_{q=1}^{n-1} \frac{1}{\La^3(w_q)}\ .
\ee

\subsection{Sector sum}

We still have to perform the sums over sector choices:
\bqa
&&\hskip-1cm\prod_{h\in L\cup\tree_L\cup E}
 \left [{\scriptstyle {4\over 3}\La^{-{1\over 2}}\lp w_{j_{h,n_h}}\rp}\right ]
\int_0^{2\pi}  d\th_{h,n_h}\;\;
\left [
{\scriptstyle{4\over 3}\La^{-{1\over 2}}\lp w_{j_{h,n_h-1}}\rp } \right ]
\int_{\Si_{j_{h,n_h}}}
d\th_{h,n_h-1}\no\\
&&...\;\;
\left [{\scriptstyle {4\over 3}\La^{-{1\over 2}}\lp w_{j_{h,1}}\rp}\right ]
\int_{\Si_{j_{h,2}}}
d\th_{h,1}
\no\\
&&
\hskip-1cm\prod_{g_i|\ i=r\ {\rm or} \atop |eg_i({\cal C})|\leq 10}
 \Upsilon\lp\th_i^{root}\{\th_{h,r(i)}\}_{h\in eg^\ast_i}\rp
 \prod_{v}
 \Upsilon\lp\th_{h_v^{root}},
\{\th_{h,n_h}\}_{ h\in H^\ast(v)}  \rp\ , \no\\ \label{secsum}
\eqa
where the products
$\left [\prod_{h\in \tree_L\cup L\cup E}\prod_{r=2}^{n_{h}}
\chi^{\th_{h,1}}_{\al_{j_{h,r}}}(\th_{h,r})
\right ]$
have been bounded by one.

We perform the sums for each half-line starting from the lowest scale
$i(h)$  and going up towards the leaves (that means the vertices).
The sum over the  root sector is bounded  by $\La^{-{1\over 2}}_T$ .
The sums for different half-lines are mixed
by the $\Upsilon$ function.

For any band $i$  we consider the
subgraph $g_i$. If $|eg_i({\cal C})|\geq 11$ and $i\neq r$
there is no $\Upsilon$ function for
this subgraph and only lines with $i(h)={\cal A}(i)$  are refined.
Hence we have to perform
\be
\prod_{h\in eg^\ast_i({\cal C})
\atop j_{h,1}=i(h)={\cal A}(i)}\hskip-.3cm\left \{
\left [{\scriptstyle {4\over 3}\La^{-{1\over 2}}\lp w_{j_{h,1}}\rp}\right ]
\int_{\Si_{j_{h,2}}}
d\th_{h,1}\right \} 1
\leq K^{\#
\left \{{h\in  eg^\ast_i({\cal C})
\atop j_{h,1}=i(h)={\cal A}(i)}\right \}}\hskip-.5cm
\prod_{h\in eg^\ast_i({\cal C})\atop j_{h,1}=i(h)={\cal A}(i)}
\frac{ \La^{{1\over 2}}(w_{j_{h,2}})}{\La^{{1\over 2}}(w_{j_{h,1}})}
\ee

If $| eg_i({\cal C})|\leq  10$, or $i=r$
we have an $\Upsilon$ function expressing
the momentum conservation at  this subgraph, and all external fields
have been refined. Each field $h\in eg_i$ except $h_i^{root}$ is refined at 
the scale ${\cal A}(i)=j_{h,r(i)}$. Hence we have to perform
\be
\prod_{h\in eg^\ast_i({\cal C})}
\left [{\scriptstyle {4\over 3}\La^{-{1\over 2}}\lp w_{j_{h,r(i)}}\rp}\right ]
\int_{\Si_{j_{h,r(i)+1}}}
d\th_{h,r(i)}
\Upsilon\lp\th_i^{root},\{\th_{h,r(i)}\}_{h\in
eg^\ast_i({\cal C})} \rp \ .\label{sec4}
\ee
We know that the function $\Upsilon$ reduces the size of the integrals
to perform. Actually  we can apply Lemma \ref{sectorlemma} below, 
which states that once the sectors for $|eg_i({\cal C})|-2$
external lines have been fixed, the last two sectors are
automatically fixed. This means that, since the sector $\th_i^{root}$ is
always fixed, we have to perform the sector sum only for
$|eg_i({\cal C})|-3$ external lines.

\begin{lemma}
Let  $\Si_i:= (\al^{-1/4},\th_i^s)$ for
$i=1,..l$ be a set of $l\geq 2$
sectors on the Fermi surface centered on $\th_i^s$ of size $\al^{-1/4}$.
Let the sector center  $\th^s_1$
be fixed, and the other sector centers $\th_i^s$ vary over intervals
$\Om_i$ of the Fermi surface:  $\th^s_i\in \Om_i$, for $i=2,..l$.
We assume $|\Om_i|>\al^{-1/4}$.
We define the function $\Upsilon(\{\th_i^s\})$ to be zero, unless
there exist some set of momenta
$\vec{k}_1,...\vec{k}_l$  satisfying
\[
\sum_{i=1}^l \vec{k}_i=0, \quad ; \quad
| |\vec{k}_i|- 1|\leq 1/\sqrt{\al}\quad \forall i \quad ; \quad
\vec{k}_i\in\Si_i \quad \forall i,
\]
($\vec{k}_i\in\Si_i $ in radial coordinates
means $|\th_i-\th^s_i|\leq \al^{-1/4}$).

Then the integral over $\th^s_{i}\in \Om_i$ of
the $\Upsilon$ constraint is bounded by
\be
\prod_{i=2}^l\left \{
\left [{\scriptstyle {4\over 3} \al^{1\over 4}}\right ]
\int_{\Om_i}
d\th_i^s \right \}
\Upsilon(\{\th^s_{i}\}_{i=1,..l} )
\leq
K^l \prod_{i\in I} \frac{|\Om_i |}{\al^{-{1\over 4}}}
\ee
where $I$ is the subset of indices of the
$l-3$ largest intervals among $\Om_2,...\Om_l$, if $l\geq 4$, and
$I=\emptyset$ if $l=2$.
\label{sectorlemma}
\end{lemma}

\paragraph{Proof}
The proof when $l\geq 4$ 
is almost identical to the one of {\bf Lemma} 3' in [FMRT1],
but we include it in Appendix B for completeness.
For $l=2$ the proof is a direct consequence of  impulsion conservation.
\qed

With these results we can bound the sum (\ref{sec4}) by
\be
K^{|eg_i|}
\prod_{h\in I(i)}\lp \frac{\La^{1\over 2}\lp w_{j_{h,r(i)+1}}\rp }
{\La^{1\over 2}\lp w_{j_{h,r(i)}}\rp}\rp
\ee
where we define $I(i)$ as the set of $|eg_i({\cal C})|-3$ half-lines
$h\in eg_i$, different from $h_i^{root}$, that have the largest
sectors $\Si_{j_{h,r(i)+1}}$.
For the particular case of $g_r$ we have the bound
\be
K^{2p}
\prod_{e\in I(1)}\lp \frac{\La^{1\over 2}\lp w_{j_{e,r(1)+1}}\rp }
{\La^{1\over 2}\lp w_{j_{e,r(1)}}\rp}\rp
\ee

We still have to consider the sums over the largest sectors: they correspond
to the vertices. Each vertex $v\in V$
can be treated as a subgraph with $|eg|\leq 10$, hence we can
apply lemma \ref{sectorlemma} with $\Om_i=[0,2\pi]$ $\forall i$,
and obtain:
\bqa
&&\prod_{v\in V} \prod_{h\in H^\ast(v)}\left \{
\left [{\scriptstyle {4\over 3}\La^{-{1\over 2}}\lp w_{j_{h,n_h}}\rp}\right ]
\int_{0}^{2\pi}
d\th_{h,n_h}\right \}
 \Upsilon\lp\th_v^{root},\{\th_{h,n_h}\}_{h\in H^\ast(v)} \rp \no\\
&&\leq
K^{4}
\La^{-{1\over 2}}\lp w_{i_v}\rp 
 \label{sec5}
\eqa
(where inessential constants such as $|\Omega_i|=2\pi $
are absorbed in a redefinition of $K$).

Remark that the refinement operations and the counting lemmas,
also cost some constants. Hence we must check that:
\begin{lemma}
The refinement and counting operations for tree and loop half-lines
altogether at most cost $K^{n}$ for some constant $K$.
\end{lemma}
\paragraph{Proof}
At each band $b=i$ with $i\geq 1$ we consider the subgraph $g_i$
(there is just one per band).

If $|eg_i({\cal C})|\leq 10 $  or $i=r$ we refine all external fields
(tree, loop and real external), and we get a factor
$K^{|eg_i({\cal C})|} \leq K^{10}$.

If $|eg_i({\cal C})|\geq 11 $ and $i\neq r$ we just refine  fields
with $i(h)={\cal A}(i)$ (there is no external field with $i(e)>0$, hence they
are are never refined in this case).
On the whole we have to pay at most

\be
\biggr( \prod_{ g_i| i=r \ {\rm or} \atop |eg_i({\cal C})|\leq 10 }
K^{10} \biggr)  K^{4n}  \ ,
\;\;
\ee
where the last factor comes from the finest refinement for each internal
and external field (there are at most $4n$ such fields). 
Now $\#\{g_i: \  |eg_i({\cal C})|\leq 10\}
\leq n\;$. This ends the proof.
\qed

\subsection{Main bound}

With all these elements, we can bound the sum (\ref{conv2}):
\bqa
\lefteqn{|\Ga_{2p\; {\rm conv.}}^{\La\Lazero}| \leq 
e^{-a\;(1-\vep)\;\La_T^{1\over s}\;  d^{1\over s}_T(\Om_1,...,\Om_{2p}) }
}\no\\
&&\hskip-.4cm K_0\;
{\scriptstyle||\phi_1||_{_1}
\prod_{i=2}^{2p} ||\hat{\phi}_i||_\infty }
\sum_{n=1}^\infty \frac{c^n}{n!} K^{n}
\sum_{CTS}\sum_{u-\tree}\sum_{\cal L}\sum_{\Om\; E}
\sum_{{\cal C}_{\rm c}}
\int_{w_T\le w_{{\cal A}(i)}
\le  w_{i}\le 1} \prod_{i=1}^{n-1} dw_i \no\\
&& \left [ \La^{5\over 2}(w_T)\right ]^{(2p-1)}
\prod_{i=1}^{n-1} \frac{1}{\La^3(w_i)}
 \prod_{a\in L}
\;\La^{3\over 4}\lp w_{M(a,{\cal C})}\rp
 \left [1
-{\La\lp {w}_{{\cal A}(m(a,{\cal C}))}\rp\over \La\lp {w}_{M(a,{\cal C})}\rp
}\right ]^{1\over 2}\no\\
&&
\prod_{g_i|\ i\neq r\ {\rm or} \atop |eg_i({\cal C})|\geq 11}\left [
\prod_{h\in eg^\ast_i \atop j_{h,1}=i(h)={\cal A}(i)}
\frac{ \La^{1\over 2}\lp w_{j_{h,2}}\rp}{\La^{1\over 2}\lp w_{j_{h,1}}\rp }
\right ]
\prod_{g_i|\ i=r\ {\rm or} \atop |eg_i({\cal C})|\leq 10 }
 \left [
\prod_{h\in I(i)} \frac{\La^{1\over 2}\lp w_{j_{h,r(i)+1}}\rp}
{\La^{1\over 2}\lp w_{j_{h,r(i)}}\rp}
\right ] \no\\
&&
 \La_T^{-{1\over 2}}\;
\prod_{v\in V}\La^{-{1\over 2}}\lp w_{i_v}\rp
\eqa
where we have bounded $|\la|\leq c$. 
Now we can send $\La$ to zero, hence $\La(w)=\sqrt{w}$ as $\La_0=1$.
The equation becomes
\bqa
&&\hspace{-0.8cm} |\Ga_{2p\; {\rm conv.}}^{\Lazero}|\leq
K_0\;
e^{-a\;(1-\vep)\;\La_T^{1\over s}\;  d^{1\over s}_T(\Om_1,...,\Om_{2p}) }
{\scriptstyle ||\phi_1||_{_1}
 \prod_{i=2}^{2p} ||\hat{\phi}_i||_\infty }
\sum_{n=1}^\infty \frac{c^n}{n!} K^{n} \\
&&\sum_{CTS}\sum_{u-\tree}\sum_{\cal L}\sum_{\Om\; E}
\sum_{{\cal C}_{\rm c}}
\ \int_{w_T\le w_{{\cal A}(i)}
\le w_{i}\le 1}
\prod_{i=1}^{n-1} dw_i
\;
\prod_{i=1}^{n-1} w_i^{-{3\over 2}} \
 \prod_{a\in L}
\; w_{M(a,{\cal C})}^{3\over 8}\;
\prod_{v\in V }w_{i_v}^{-{1\over 4}}
\no\\ &&
\prod_{g_i|\ i\neq r \ {\rm or}\atop |eg_i({\cal C})|\geq 11}\left [
\prod_{h\in et_i^\ast\cup el_i\atop j_{h,1}=i(h)={\cal A}(i)}
\frac{w_{j_{h,2}}^{1\over 4}}{w_{j_{h,1}}^{1\over 4}}\right ]
\prod_{g_i|\ i=r \ {\rm or} \atop |eg_i({\cal C})|\leq 10 }
 \left [
\prod_{h\in I(i)} \frac{w_{j_{h,r(i)+1}}^{1\over 4}}
{w_{j_{h,r(i)}}^{1\over 4}}
\right ] \;\; w_T^{{5p\over 2}-{3\over 2}}  \no
\eqa
where we have bounded
$\left [1
-{\La({w}_{{\cal A}(m(a,{\cal C}))})\over \La({w}_{M(a,{\cal C})})
}\right ]^{\frac{1}{2}}$ by one.

To factorize the integrals we perform the  change of variable:
\be
w_i= \frac{1}{\bt_i} w_{{\cal A}(i)} \qquad  1\leq i\leq n-1.
\label{fact1}\ee
By  (\ref{bornew}) we have the following bound for $\beta_i$
\be
\beta_i \in
\left [ \frac{ w_{{\cal A}(i)}}{\min[w_{i'},w_{i''}]},1 \right ]
\label{fact2}\ee
Now each $w_i$ can be written
\be
w_i = \left [\prod_{j\in C_i}  \frac{1}{\bt_j} \right ] w_T
\label{fact3}\ee
where we defined $C_i$ as the set of crosses on the chain joining
the cross $i$ to the root.
The Jacobian of this transformation is the determinant of
the matrix
\[
M_{ij}= \frac{\partial w_i}{\partial \beta_j} =  - \frac{1}{\bt_j} w_i\;
\chi(j\in C_{i}) \ 
\]
where $\chi(j\in C_{i})=1$ if $j\in C_{i}$ and $0$ otherwise.
If we order the rows and columns  of $M_{ij}$ putting the root first, then the first layer of the $CTS$ and so on, we see that $M_{ij}$ is a
triangular matrix, hence its determinant is given by:
\be
|Jac| =\left | \prod_{i=1}^{n-1} \frac{\partial w_i}{\partial \beta_i}
\right |  =  w_T^{n-1}  \prod_{i=1}^{n-1}\left [\frac{1}{\bt_i}
 \prod_{j\in C_i} \frac{1}{\bt_j}\right ] =
w_T^{n-1}  \prod_{i=1}^{n-1}\left [\frac{1}{\bt_i}  \lp\frac{1}{\bt_i}
\rp^{n_i-1} \right ]
\ee
where $n_i$ is the number of vertices in the subgraph $g_i$. Indeed
$\beta_i$ appears in the chain $C_j$ exactly for all $j\geq_P i$, hence 
its exponent is the number of crosses above $i$, which is the number of 
tree lines in $g_i$, hence $n_{i}-1$ if we denote the number
of vertices in $g_i$ by $n_i$. In these new coordinates we have:
\bqa
&&\hspace{-0.8cm} |\Ga_{2p\; {\rm conv.}}^{\Lazero}|\leq K_0\;
{\scriptstyle ||\phi_1||_{_1}
 \prod_{i=2}^{2p} ||\hat{\phi}_i||_\infty }
e^{-a\;(1-\vep)\;\La_T^{1\over s}\;  d^{1\over s}_T(\Om_1,...,\Om_{2p}) }
\sum_{n=1}^\infty \frac{c^n}{n!} K^{n}\no\\
&&\sum_{CTS}\sum_{u-\tree}\sum_{\cal L}\sum_{\Om\; E}
\sum_{{\cal C}_{\rm c}}
\int_{w_T}^1 \prod_{i=1}^{n-1} d\bt_i
 \; w_T^{n-1} \prod_{i=1}^{ n-1} \bt_i^{-1+(1-n_{i})}
\prod_{i=1}^{n-1}\left [
\lp\prod_{j\in C_i}\bt_j^{3\over 2}\rp w_T^{-{3\over 2}}\right ]\no\\
&&\prod_{a\in L}\left [
\;\lp\prod_{j\in C_{M(a,{\cal C})}}\bt_{j}^{-{3\over 8}}\rp w_T^{3\over 8}
\right ]
\; \prod_{g_i|\ i\neq r\ {\rm or}\atop |eg_i({\cal C})|\geq 11}\left [
\prod_{h\in eg_i^\ast\atop j_{h,1}=i(h)={\cal A}(i)}
\lp \prod_{j\in C_{r(i)+1}\backslash C_{r(i)} }
\frac{1}{\bt_{j}^{1\over 4}}\rp
\right ] \no\\
&&\prod_{g_i|\ i=r \ {\rm or}\atop |eg_i({\cal C})| \leq 10 }
 \left [
\prod_{h\in I(i)}\lp\prod_{j\in C_{r(i)+1}\backslash C_{r(i)} }
\frac{1}{\bt_{j}}\rp^{1\over 4}
\right ] \;
\prod_{v \in V}\left [ \lp\prod_{j\in C_{i_v}}\bt_{j}\rp^{1\over 4}
w_T^{-{1\over 4}}\right ]
 w_T^{{5p\over 2}-{3\over 2}}\ ,\no\\
\label{gamma}\eqa
where we have taken as integration domain for all $\bt_i$ the
interval $[w_T,1]$, that contains the exact integration domain, since 
${w_{{\cal A}(i)}\over \min [w_{i'}, w_{i"}]}\geq {w_{{\cal A}(i)}\over 1}
\geq w_T$.
We write the integrals over the different $\bt_i$ as a product
$\prod_{i=1}^{n-1} \int_{w_T}^1 d\bt_i \;\bt_i^{-1+x_i}$.
We have to find out the expression for $x_i$.
We observe that
\bqa
&&\prod_{i=1}^{n-1}
\lp\prod_{j\in C_i}\bt_j^{3\over 2}\rp  =
\prod_{i=1}^{n-1}\bt_i^{{3\over 2}(n_{i}-1)}\no\\
&&\prod_{a\in L}
\lp\prod_{j\in C_{M(a,{\cal C})}}\bt_{j}^{-{3\over 8}}\rp
= \prod_{i=1}^{n-1} \bt_i^{-{3\over 8}
\; \#\{a\in L| M(a,{\cal C})\geq_P i  \} }=
\prod_{i=1}^{n-1} \bt_i^{-{3\over 8}|il_i({\cal C})|}\no\\
&&
\prod_{v\in V }
 \lp\prod_{j\in C_{i_v}}\bt_{j}\rp^{1\over 4}
=  \prod_{i=1}^{n-1} \bt_i^{{1\over 4}\# \{ v\in V| i_v\geq_P i\}} =
 \prod_{i=1}^{n-1} \bt_i^{{1\over 4}n_{i}}\ , \no
\eqa
and the remaining products over sector attributions are equal to
 $\prod_{i=1}^{n-1}   \bt_i^{-y_i}$
where
\bqa
 y_i & = & 0\quad {\rm if} \quad
|eg_i({\cal C})|=2\no\\
&=&
{1\over 4} (|eg_i({\cal C})|-3)\quad {\rm if} \quad
|eg_i({\cal C})|\leq 10\no\\
 &\le &
{1\over 4} (|eg_i({\cal C})|-1)\quad {\rm if} \quad
|eg_i({\cal C})| > 10 \no\\
 y_r & = & 0\quad {\rm if} \quad
|eg_r({\cal C})|=2\no\\
&=&
{1\over 4} (|eg_i({\cal C})|-3)\quad {\rm if} \quad
|eg_r({\cal C})|>2 \ .
 \label{ybound}
\eqa
To obtain this bound we observe that the factor $\bt_i$ appears
in the product with a power $-1/4$ each time there is a half-line
$h\in \tree_L\cup L\cup E$ with
\[ i
\in C_{r(i)+1}\backslash C_{r(i)}
\]
for some $r$ and the corresponding factor appears in the sector counting.
Now, for each subgraph $g_{i}$ we have
three situations
\begin{itemize}
\item{} $|eg_{i}({\cal C})|=2$: then the factor $\bt_i$ does not appear,
i.e. $y_{i}=0$.
\item{} $4\le |eg_{i}({\cal C})|\leq 10$, hence all external
half-lines except $h^{\rm root}$ are refined and the factor $\bt_i$
appears with power $-1/4  (|eg_{i}({\cal C})|-3)$.
\item{}  $|eg_{i}({\cal C})| > 10$: only some
of the external lines of $g_{i}$ (other than $h^{\rm root}$) 
are refined; therefore
the factor $\bt_i$ appears with power $-{1\over 4}  a_i$ where
$a_i\leq   (|eg_{i}({\cal C})|-1)$ is  the number of
 external half-lines refined. This is why (\ref{ybound})
is a bound and not an equality.
\end{itemize}

Now we can bound (\ref{gamma}) (using that $|L|=2 (n+1-p)$):
\bqa
|\Ga_{2p\; {\rm conv.}}^{\Lazero}|&\leq & K_0\;
{\scriptstyle ||\phi_1||_{_1}
\prod_{i=2}^{2p} ||\hat{\phi}_i||_\infty }
e^{-a\;(1-\vep)\;\La_T^{1\over s}\;  
d^{1\over s}_T(\Om_1,...,\Om_{2p}) } w_T^{{7p\over 4} -{1\over 4}}
\no\\ &&
\sum_{n=1}^\infty \frac{c^n}{n!} K^{n}
\sum_{CTS}\sum_{u-\tree}\sum_{\cal L}\sum_{\Om\; E}
\sum_{{\cal C}_{\rm c}}
\prod_{i=1}^{ n-1}\int_{w_T}^1  d\bt_i \; \bt_i^{-1+x_i}
\eqa
where
\be
 x_i =
 {1\over 2}(n_{i}-1) -{3\over 8} |il_i({\cal C})| +{1\over 4}n_{i}
- {1\over 4} (|eg_i({\cal C})|-3)
 \ee
when $i=r$, or
$4\le |eg_i({\cal C})|\leq 10$ and
\be
x_i \ge
{1\over 2}(n_{i}-1) -{3\over 8} |il_i({\cal C})|+{1\over 4}n_{i}
-{1\over 4} (|eg_i({\cal C})|-1))
\ee
when  $|eg_i({\cal C})| > 10$.
The integrals over $\bt_i$ are well defined only if $x_i>0\ \forall i$.
To check that it is true, we observe that
\be
{1\over 2}(n_{i}-1) -{3\over 8} |il_i({\cal C})|+{1\over 4}n_{i}
= {1\over 8} (3|eg_i({\cal C})|-10)
\ee
where we applied the relation
\[
|il_i({\cal C})| = 2 n_{i} + 2- |eg_i({\cal C})|.
\]
Hence, for  $i=r$, or
$4\le |eg_i({\cal C})|\leq 10$
we have
\be
 x_i =   {1\over 8} (3|eg_i({\cal C})|-10)
 -{1\over 4} (|eg_i({\cal C})|-3) =  {1\over 8} (|eg_i({\cal C})|- 4 )
 \label{xbound1}\ee
and when  $|eg_i({\cal C})| > 10$ (and $i\neq r$) we have
\be
 x_i \ge {1\over 8} (3|eg_i({\cal C})|-10)
 -{1\over 4} (|eg_i({\cal C})|-1) =  {1\over 8} (|eg_i({\cal C})|- 8 )\geq 
{1\over 2}
 \label{xbound2}\ee
by construction.
Remark that since the lowest subgraph $g_r$ has no tree external line we 
can compute explicitly
\be
 x_r =     {1\over 8} (|eg_r({\cal C})|- 4 ) =
 {1\over 8} (2p - 4 ) \ .
\ee
If $|eg_{i}({\cal C})| = 4$, $x_i=0$ and the graph
is logarithmic in the temperature:
\be
\int_{w_T}^1 d\bt_i \;\bt_i^{-1} = -\log w_T = 2\; \left |
\log (\sqrt{2}\pi T)\right|
\label{div1}\ee
Finally if $|eg_{i}({\cal C})| = 2$, then
\be
x_i = {1\over 8} (3|eg_{i}({\cal C})|-10) =   -{1\over 2}
\ee
and the integral over $\bt_i$ is linearly divergent with the
temperature $T$:
\be
\int_{w_T}^1 d\bt_i \;\bt_i^{-1-{1\over 2}} = 2 \;(w_T^{-{1\over 2}}-1) =
2\;\lp{1\over \sqrt{2}\pi T} -1 \rp.
\label{div2}\ee

Hence we have recovered the well known fact that the only divergent 
subgraphs are the four points and two points subgraphs [FT1-2]-[FMRT1]. 
In this paper we restrict ourselves to convergent attributions, 
for which $x_i$ is always positive. However it is important 
(in order to bound later the sum over labelings)
that we check that we have a lower bound on  $x_i$ which
is proportional to the number of external tree lines of $g_{i}$:
\begin{lemma}
For any subgraph $g_{i}$ ($i\neq r$) we have
\be
x_{i}\geq {|et_{i}| \over 72}>0 \ .
\ee
\end{lemma}
\paragraph{Proof}
We distinguish several cases:
\begin{itemize}
\item{} if $|eg_i|\leq 10$
\be
 {1\over 8}(|eg_{i}({\cal C})|-4) \geq {1\over 4}>0
\ee
as for convergent attributions $|eg_i({\cal C})|\geq 6$
 (we cannot have  $|eg_{i}({\cal C})| = 5$ by parity).
Now, if  $|et_{i}|\geq 5$ we have
\be
 {1\over 8}(|eg_{i}({\cal C})|-4)\geq {1\over 8}(|et_{i}|-4)
\geq {1\over 5\cdot 8}|et_{i}|\ .
\ee
If  $|et_{i}|\leq 4$ we can write
\be
 {1\over 8}(|eg_{i}({\cal C})| -4)\geq {1\over 4}\geq
{1\over 16}|et_{i}|\ .
\ee
\item{} if $|eg_i| > 10$  we have
\be
 {1\over 8}(|eg_{i}({\cal C})|-8) \geq {1\over 2}>0 \ .
\ee
Repeating  the same arguments as before for the case  $|et_{i}|\geq 9$  and
$|et_{i}| < 9$ we obtain
\be
 {1\over 8}(|eg_{i}({\cal C})|-8)
\geq {1\over 8\cdot 9}|et_{i}|\ .
\ee
\end{itemize}
This completes the proof of the Lemma
\qed

\begin{figure}
\centerline{\psfig{figure=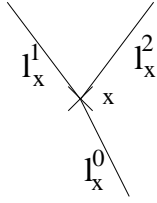,width=2cm}}
\caption{definition of $\ell^{0}_{x}$,$\ell^{1}_{x}$, $\ell^{2}_{x}$}
\label{ellinefig}
\end{figure}

Now we can perform the integrals on the $\bt_i$, to obtain
\bqa
|\Ga_{2p>4\; {\rm conv.}}^{\Lazero}| & \leq & K_0\;
{\scriptstyle ||\phi_1||_{_1}
\prod_{i=2}^{2p} ||\hat{\phi}_i||_\infty }\; 
e^{-a\;(1-\vep)\;\La_T^{1\over s}\;  d^{1\over s}_T(\Om_1,...,\Om_{2p}) }
\label{sec6}\\
&& w_T^{{7p\over 4} -{1\over 4}}
{1\over 2p-4}
\sum_{n=1}^\infty \frac{c^n}{n!} K^{n} \;
\sum_{CTS}\sum_{u-\tree}\sum_{\cal L}\sum_{\Om\; E}
\sum_{{\cal C}_{\rm c}}
\prod_{i\neq r}{ 1\over |et_i|}\ ,\no
\eqa
where the factor $\prod_i \lp 1- w_T^{x_i} \rp$ coming from the integrals
over the variables $\bt_i$ has been bounded by one.
For the particular case of four point and two point vertex functions we have
\bqa
|\Ga_{4\; {\rm conv.}}^{\Lazero}|& \leq & K_0\;
{\scriptstyle ||\phi_1||_{_1}
\prod_{i=2}^{4} ||\hat{\phi}_i||_\infty }\;
e^{-a\;(1-\vep)\;\La_T^{1\over s}\;  d^{1\over s}_T(\Om_1,...,\Om_{4}) }
\label{sec64} \\
&&  w_T^{13\over 4}|\log w_T|\;
\sum_{n=1}^\infty \frac{c^n}{n!} K^{n}
\sum_{CTS}\sum_{u-\tree}\sum_{\cal L}\sum_{\Om\; E}
\sum_{{\cal C}_{\rm c}}
\prod_{i\neq r}{ 1\over |et_i|}\ .\no
\eqa
\bqa
\hspace{-1.2cm}|\Ga_{2\; {\rm conv.}}^{\Lazero}| &\leq & K_0\;
{\scriptstyle ||\phi_1||_{_1}
 ||\hat{\phi}_2||_\infty } \;
e^{-a\;(1-\vep)\;\La_T^{1\over s}\;  d^{1\over s}_T(\Om_1,\Om_{2}) }
\label{sec642}\\
&&  w_T 
\sum_{n=1}^\infty \frac{c^n}{n!} K^{n}
\;
\sum_{CTS}\sum_{u-\tree}\sum_{\cal L}\sum_{\Om\; E}
\sum_{{\cal C}_{\rm c}}
\prod_{i\neq r}{ 1\over |et_i|}\ .\no
\eqa
The sum $\sum_{{\cal C}_{\rm c}}$ is over a
set whose cardinal is bounded
by  $K^{n}$ so we can  bound it with the supremum
over the set. The sum over
$\Om $ runs over a set of at most $2^{n-1}$ elements.
The sum over $E$ to attribute the $2p$ external lines to particular
vertices runs over a set of at most $ n^{2p}$ (this is an overestimate!).
Hence 
\[
\sum_{{\cal C}_{\rm c}}\sum_{\Om}\sum_E | F ({\cal C}_{\rm c}, \Om, E)|
\le (p!)^2 K^n \ \sup_{{\cal C}_{\rm c}, \Om, E}
| F ({\cal C}_{\rm c}, \Om, E)| \ ,
\]
where we applied the bound 
\[
n^{2p} \le  (2p)! e^n \leq K^p\ (p!)^2 e^n \quad \forall n\geq 0\ .
\]
We still have to perform the sum over the $CTS$ and ${\cal L}$.
For each cross $x$ of the $CTS$ different from the
root, there is one line $\ell^{0}_{x}$ going down (towards the root), and
two lines $\ell^{1}_{x}$ and $\ell^{2}_{x}$ going up (see Fig.\ref{ellinefig}).

\begin{lemma}
For any cross $x$ different from the root~:
\be N_{\ell^{0}_{x}}(\tree,{\cal L})= N_{\ell^{1}_{x}}(\tree,{\cal L}) +
N_{\ell^{2}_{x}}(\tree,{\cal L}) -2
\ee
\end{lemma}
\noindent{\bf Proof}: The clusters $\tree_{\ell^{1}_{x}}(\cal L)$ and
$\tree_{\ell^{2}_{x}}(\cal L)$ are joined by a single line in the tree $\tree$,
which is the label of the cross $x$. This line is counted once
as external line of $\tree_{\ell^{1}_{x}}(\cal L)$ and once as external line of
$\tree _{\ell^{2}_{x}}(\cal L)$, and is no longer an external line of
$\tree _{\ell^{0}_{x}}(\cal L)$. This proves the lemma.

\qed

The following Lemma is an improved version of Lemmas B4-B5 in [CR] (see also
Lemma III.6), adapted to this formalism of relative rather than total
orderings.
\begin{lemma}
Let $CTS$ be a fixed Clustering
Tree Structure of order $n$. We have \be
\sum_{\tree} \sum_{{\cal L}} {1\over n~!}
\prod_{\ell} {1 \over N_{\ell}(\tree,{\cal L})}
\le 4^{n}\ .
\ee
\end{lemma}
\noindent{\bf Proof}: We decompose the sum over $\tree$ and ${\cal L}$ into
subsums. We call ${\cal L}_{o}$ the map which associates the dots of $CTS$
to the vertices of $\tree$ and  ${\cal L}_{x}$ the map which associates the
crosses of $CTS$ to the lines of $\tree$. By the previous lemma,
once ${\cal L}_{o}$ and the collection $V=\{N_{v}\}$ of coordination numbers
for each vertex $v$ of $\tree$ is given, the numbers $N_{\ell}(\tree,{\cal L})$
are all fixed,
hence they do not depend on the particular contractions $W$ and on
${\cal L}_{x}$.
This suggests to split the sum over $\tree$ and ${\cal L}$ as a sum
over  $W$ and ${\cal L}_{x}$ followed by a sum over $V$ and ${\cal L}_{o}$~:
\be
\sum_{\tree} \sum_{{\cal L}} {1\over n!} \prod_{\ell}
{1 \over N_{\ell}(\tree,{\cal L})} =
\sum_{V}\sum_{{\cal L}_{o}}{1\over n!} \prod_{\ell} {1 \over
N_{\ell}(V,{\cal L}_{o})}
\sum_{{\cal L}_{x}} \sum_{W} 1\ .
\ee
But the number of labelings ${\cal L}_{x}$ and contractions $W$ compatible
with given $V$ and ${\cal L}_{o}$ is precisely $\prod_{\ell}
N_{\ell}(V,{\cal L}_{o}) $. Indeed starting from the $n$ dots in $CTS$ with 
their $N_{v}$ hooked fields, and going down towards the root
we can inductively build the contractions
corresponding to each cross of $CTS$ (this builds at the same time
$W$ and ${\cal L}_{x}$). To count the possible contractions for a cross $x$,
we have to choose one external field in  $T_{\ell^{1}_{x}}$ and one
in  $T_{\ell^{2}_{x}}$, hence the number of choices is  {\it exactly}
$N_{\ell^{1}_{x}}(V,{\cal L}_{o})N_{\ell^{2}_{x}}(V,{\cal L}_{o})$, where
$\ell^{1}_{x}$ and $\ell^{2}_{x}$ were introduced in the previous
lemma. Multiplying over all crosses, we get~:
\be
\sum_{\tree} \sum_{\cal L} {1\over n~!} \prod_{\ell}
{1 \over N_{\ell}(V,{\cal L}_{o})} =
\sum_{V}\sum_{{\cal L}_{o}}{1\over n~!} = \sum_{V} 1 \le 4^{n}\ .
\ee
Indeed $n!$ is exactly the number of
labelings ${\cal L}_{o}$ of the dots of $CTS$, and for each vertex $v$
$N_{v}$ is an integer between 1 and 4, hence the sum over $V$ is bounded
by $4^{n}$ (this is an upper bound since we do not take into account
the constraint $\sum_{v} N_{v} = 2n-2$).
\qed

Applying the lemma above, we bound
\be
\frac{1}{n!}
\sum_{u-\tree}\sum_{\cal L}
\prod_{i\neq r}{ 1\over |et_i|}
\leq 4^n\ .
\ee
Hence the vertex function is bounded by
\bqa
&&\hspace{-4cm}|\Ga_{2p>4\; {\rm conv.}}^{\Lazero}|\;\leq \;  K_0\;
{\scriptstyle ||\phi_1||_{_1}
\prod_{i=2}^{2p} ||\hat{\phi}_i||_\infty } 
e^{-a\;(1-\vep)\;\La_T^{1\over s}\;  d^{1\over s}_T(\Om_1,...,\Om_{2p}) }\no\\
&&{w_T^{{7p\over 4} -{1\over 4}}\over 2p-4} \ K_1^p\  \lp p!\rp^2 
\sum_{n=1}^\infty c^n K_2^{n}
\eqa
\be
|\Ga_{4\; {\rm conv.}}^{\Lazero}|\leq  K'_0\;
{\scriptstyle ||\phi_1||_{_1}
\prod_{i=2}^{4} ||\hat{\phi}_i||_\infty }\;
e^{-a\;(1-\vep)\;\La_T^{1\over s}\;  d^{1\over s}_T(\Om_1,...,\Om_{4}) }\;
w_T^{13\over 4}  \; |\log w_T|\; 
\sum_{n=1}^\infty c^n K_2^{n}
\ee
\be
|\Ga_{2\; {\rm conv.}}^{\Lazero}|\leq  K''_0\;
{\scriptstyle ||\phi_1||_{_1}
 ||\hat{\phi}_2||_\infty }\;
e^{-a\;(1-\vep)\;\La_T^{1\over s}\;  d^{1\over s}_T(\Om_1,\Om_{2}) }
\; w_T \ 
\sum_{n=1}^\infty c^n K_2^{n}
\ee
for some constant $K_2$. 
This is convergent for $c < {1\over K_2}$ 
and achieves the proof of Theorem 1 and 2. (Remark that we did not try to 
optimize the dependence of this bound in $2p$, the number of external points).
\newpage

\resetsect 

\renewcommand{\thesection}{\Alph{section}}

\noindent{\Large {\bf Appendix A }}
\medskip

\noindent{\large {\bf Spatial decay}}
\vskip 0.5cm

\resetequ

We prove that the $T=0$ propagator $C_0$ decays as
\bqa
\lefteqn{| C_0^{w_q}
( x,0, \th_{h_q,1})|
\leq}\label{spdec1}\\
&& K \; \frac{\La_0^2-\La^2}{\La^4(w_q)}
\;\La^{{1\over 2}}(w_q) \;\La^3(w_q)\; e^{-a\left [
|x_0 \La(w_q)|^{{1\over s}}+|x_r \La(w_q)|^{{1\over s}}+
 |x_t \La^{{1\over 2}}(w_q)|^{{1\over s}}
\right ]}\ .
\no\eqa

\begin{lemma}
Let $f\in {\cal C}^\infty({\rm \RR}^d)$ be such that its Fourier transform
$\hat{f}$ has compact support of volume $V_f$ and satisfies
\be
||\hat{f}^{(n_1,...,n_d)}||_\infty :=
\left | \left |\frac{\partial^{n_1}}{\partial p_1^{n_1}}...
\frac{\partial^{n_d}}{\partial p_d^{n_d}} \hat{f}\right |\right |_\infty
\leq A_0 \prod_{i=1}^d \left [
\lp\al_i C\rp^{n_i}\; (n_i!)^s\right ],
\ee
where $A_0$, $C$, $\al_1$,...$\al_d$
are some constants and $s\geq 1$ is some constant.

Then for some constants $K$, $\mu$  and $a$, one has
\be
 |f(x)| \leq
K\; A_0\; V_f \; e^{-a\sum_{i=1}^d \left |{x_i\over \al_i} 
\right |^{1/s}}\quad
\forall x\in {\rm \RR}^d  \ .
\ee
\end{lemma}
\paragraph{Proof.}
By Stirling's formula the first equation can be written
\be
||\hat{f}^{(n_1,...,n_d)}||_\infty\leq
 A_0\; K\; \prod_{i=1}^d \left [
 \lp{\al_i\over \mu}\rp^{n_i}\; \lp{n_i\over e}\rp^{n_i s}\right ]
\ee
where $K$ and $\mu$ are some constants (eventually dependent from $d$).
Hence, for any $x$ we have:
\bqa
\hskip-.7cm |f(x)| &=& \left | {1\over (ix_1)^{n_1}...(ix_d)^{n_d}}
\int e^{-ipx} \; \hat{f}^{(n_1,...,n_d)}(p) \right |\no\\
&\leq & {||\hat{f}^{(n_1,...,n_d)}||_\infty \; V_f
\over |x_1|^{n_1}...|x_d|^{n_d}} \leq V_f\;K\; A_0\;
\prod_{i=1}^d  \left [
 \left |{\al_i\over \mu x_i}\right |^{n_i}\; \lp{n_i\over e}\rp^{n_i s}\right ]
\ .
\eqa
Optimizing to $n_i=
\left |{\mu x_i\over \al_i}\right |^{1\over s}$, we obtain
\be
|f(x)| \leq V_f\; A_0\;K\;
\prod_{i=1}^d  e^{-s
 \left |{\mu x_i\over \al_i}\right |^{1\over s}}
\ee
which ends the proof of the lemma, with $a=s\mu^{1\over s}$.
\qed

\begin{lemma}
$C_0^{w_q}$ satisfies (\ref{spdec1}).
\end{lemma}
\paragraph{Proof}
To prove (\ref{spdec1}) we write in momentum space:
\be
| C_0^{w_q}
( k_r,k_t, \th_{h_q,1})|\hskip-.05cm=\hskip-.05cm
\left |C_0^{w_{q}}(k)
\chi^{\th_{h_q,1}}_{\al_{j_{h_q,1}}}[\th(k_r,k_t)]
\right |= \hskip-.1cm\frac{(\La_0^2-\La^2)}{\La^4(w_q)}
\left |C_0^{w_{q},\th_{h_q,1}}(k_0,k_r,k_t)\right |
\ee
where the radial and tangential variables $k_r$ and $k_t$
are defined by:
\be
k_r =  |\vec{k}| \cos (\th-\th_{h_q,1})-1 \quad ; \quad
k_t =  |\vec{k}| \sin (\th-\th_{h_q,1}) \ .
\ee
The function to study is (since $j_{h_q,1}=q$):
\be
C_0^{w_{q},\th_{h_q,1}}(k_0,k_r,k_t)=
u_p\left[\al^{1/4}_{q}(\th-\th_{h_q,1})\right ]
\;\;
[ik_0+e(|\vec{k}|)] \; u'[\al_q(k_0^2+e^2(|\vec{k}|)],
\ee
and
\bqa
\th-\th_{h_q,1} &=& f_1(k_r,k_t)= \arctan {k_t\over 1+ k_r}\no\\
e(|\vec{k}|)&=&  |\vec{k}|^2-1= f_2(k_r,k_t) = k_r^2 + k_t^2 +2 k_r
\label{gv1}
\eqa
The propagator $C_0^{w_{q},\th_{h_q,1}}$ can be written  as 
the product of three 
functions
\be
C_0^{w_{q},\th_{h_q,1}}(k_0,k_r,k_t) = F_1(k_r,k_t) \ F_2(k_0,k_r,k_t)
 \ F_3(k_0,k_r,k_t)
\ee
where
\bqa
F_1(k_r,k_t)&:=& u_p\lp\al_{q}^{1\over 4} f_1(k_r,k_t)\rp
\no\\
F_2(k_0,k_r,k_t)&:=& [ik_0+ f_2(k_r,k_t) )] \no\\
F_3(k_0,k_r,k_t)&:=&
u'[\al_{q}(k_0^2+f_2^2(k_r,k_t))] \ ,
\label{gv2}
\eqa
 $f_1$, $f_2$ being defined in  (\ref{gv1}).

Now we know that $u(x)$ is a Gevrey function of class $s$,
with compact support on $\left [-{1\over 2},{1\over 2}\right ]$.
The function $f_1$ takes values in the interval
$\left[-{\al_{q}^{-1/4}\over 2},{\al_{q}^{-1/4}\over 2}\right]$, hence
$k_r\in \left [-{\al_{q}^{-{1\over 2}}\over 2},
{\al_{q}^{-{1\over 2}}\over 2}\right]$,
$k_t\in \left [-{\al_{q}^{-1/4}\over 2},{\al_{q}^{-1/4}\over 2}\right]$.
By hand or using the standard
rules for derivation, product and composition of Gevrey functions (see [G])
it is then easy to check that
$C_0^{w_q,\th_{h_q,1}}(k_0,k_r,k_t)$ is a Gevrey function
with compact support of
class $s$ and satisfies the bound:
\be
\left |\left |\frac{\partial^{n_0}}{\partial k_0^{n_0}}
\frac{\partial^{n_r}}{\partial k_r^{n_r}}
\frac{\partial^{n_t}}{\partial k_t^{n_t}} C_0^{w_q,\th_s}
\right |\right |_\infty
\leq  { 1\over{\sqrt{\al_{q}}}}
C_0^{n_0+n_t+n_t}\; \lp\al_q^{1\over 2}\rp^{n_r+n_0}  \;
\lp \al_q^{1\over 4}\rp^{n_t}\;
 \lp n_0! n_r! n_t!\rp^s \, .
\ee
Hence, applying  Lemma 2, with $A_0=  1/\al_q^{{1\over 2}}$
and $V_f= \La^{{1\over 2}}(w_q) \La^2(w_q)$, proves
(\ref{spdec1}).
\qed
\vskip 1cm

\setcounter{section}{2}

\renewcommand{\thesection}{\Alph{section}}

\noindent{\Large {\bf Appendix B }}
\medskip

\noindent{\large {\bf Proof of the Sector Counting Lemma \ref{sectorlemma}  }}
\vskip 0.5cm

\resetequ

We define $\vec{k}'_i$ as the projection of $\vec{k}_i$ on the
Fermi surface $\vec{k}'_i=\vec{k}_i/|\vec{k}_i|$ and $\vec{r}_i$
as the center of the sector $\Si_i$, with components $(1,\th^s_i)$ in radial
coordinates. Then, as in [FMRT1], we renumber $\vec{k}_2,...\vec{k}_l$ so that
$|\vec{r}_l \cdot \vec{r}_{l-1}|$ is the minimum of the set
$\{|\vec{r}_i\cdot\vec{r}_j|| i,j>1 \}$. This means that the angle between
$\vec{k}'_l$ and  $\vec{k}'_{l-1}$
$\phi:= \angle (\vec{k}'_{l-1}, \vec{k}'_{l} )$
is as close as possible to $\pi/2$.
All other angles $ \angle (\vec{k}'_{i}, \vec{k}'_{j} )$ with $i,j\geq 2$
must be within  $\phi+O(\al^{-1/4})$ of either 0 or $\pi$.
The proof is performed in two steps.
\paragraph{1.} When $2^{-i}\leq |\phi| \leq 2^{-i+1}$  or
 $2^{-i}\leq |\pi-\phi| \leq 2^{-i+1}$, for any $i$ fixed, we have
\be
N_l := \left [{\scriptstyle {4\over 3} \al^{1\over 4}}\right ]^2
\int_{\Om_l}  d\th_l^s
\int_{\Om_{l-1}}  d\th_{l-1}^s
\Upsilon(\{\th^s_{i}\}_{i=1,..l} )
\leq K_0^l \; \left [
{\scriptstyle {4\over 3} \al^{1\over 4}}\right ]^2 \lp
\al^{-{1\over 4}}\rp^2\leq
K^l
\label{bound}\ee
where $K_0$ and $K$ are some constants and  the sector centers
$\th_2,...\th_{l-2}$, are not integrated yet.
The proof is shown below.
\paragraph{2.}
We now have to perform the remaining integrals, then sum over all
possible values of $i$.
Assuming (\ref{bound})  true, the sum over all sectors is bounded by
\bqa
\prod_{j=2}^l
\left [{\scriptstyle {4\over 3} \al^{1\over 4}}\right ]
\int_{\Om_j}
d\th_j^s \;
\Upsilon(\{\th^s_{j}\}_{j=1,..l} )
&\leq &
K^l  \prod_{j\in J(i)}
\left [{\scriptstyle {4\over 3} \al^{1\over 4}}\right ]
\int_{2^{-i}}
d\th_j^s
\prod_{j\not\in J(i)}
\left [{\scriptstyle {4\over 3} \al^{1\over 4}}\right ]
\int_{\Om_j}
d\th_j^s \, 1\no\\
&= & K'^l \;
 \prod_{j\in J(i)} \frac{2^{-i}}{\al^{-{1\over 4}}}
 \prod_{j\not\in J(i)} \frac{|\Om_j  |}{\al^{-{1\over 4}}}
 \eqa
where
$J(i):= \{j | 2^{-i}\leq |\Om_j|, 1<j<l-1\}$.
To perform the sum over all possible $i$ we distinguish two situations,
defining $i_{0}$ such that $2^{-i_0} \le |\Om_l | < 2^{-i_0+1}$:
\begin{itemize}
\item{}if $2^{-i}\leq 2^{-i_0}$,
we have to perform
\be\hskip-1cm
\sum_{i=i_0}^\infty  \prod_{j\in J(i)} \frac{2^{-i}}{\al^{-{1\over 4}}}
 \prod_{j\not\in J(i)} \frac{|\Om_j |}{\al^{-{1\over 4}}}
\leq \sum_{i=i_0}^\infty
 \frac{2^{-i}}{\al^{-{1\over 4}}} \prod_{j=3}^{l-2}
\frac{|\Om_j |}{\al^{-{1\over 4}}}
=  \frac{2^{-i_{0}+1}}{\al^{-{1\over 4}}}  \prod_{j=3}^{l-2}
\frac{|\Om_j |}{\al^{-{1\over 4}}}
 \leq
2 \prod_{i\in I} \frac{|\Om_i|}{\al^{-{1\over 4}}} \ .
\ee
\item{}if $2^{-i}> 2^{-i_0}$, then, once fixed the sectors of all the
$\vec{k}_i$ except the $l^{th}$ there can be at most one $i$ consistent with
$\vec{k}_l$ falling in $\Om_l$. For this single value of $i$
\be
\prod_{j\in J(i)} \frac{2^{-i}}{\al^{-{1\over 4}}}
\prod_{j\not\in J(i)} \frac{|\Om_j|}{\al^{-{1\over 4}}}
\leq  \frac{\prod_{j=2}^{l-2}|\Om_j|}{\al^{-{l-3\over 4}}}\leq
\prod_{i\in I} \frac{|\Om_i|}{\al^{-{1\over 4}}}\ .
\ee
\end{itemize}
This completes step 2 of the proof. We perform now step 1.
\paragraph{Proof of (\ref{bound})} (almost identical to [FMRT1], pg 701-704).
We introduce the vectors $\vec{a}$ and $\vec{\vep}$ defined as
\bqa
\vec{a} &=& -\vec{r}_1 - ... - \vec{r}_{l-2}\; =\;
\vec{k}_{l-1} + \vec{k}_l -
\sum_{j=1}^{l-2} (\vec{r}_j-\vec{k}_j)\no\\
\vec{a}+\vec{\vep} &=& \vec{k}'_l +  \vec{k}'_{l-1}
    = -\vec{k}_1 - ... - \vec{k}_{l-2} + 2 O(\al^{-{1\over 2}})
\eqa
hence
\be
\vec{\vep} =
\sum_{j=1}^{l-2} (\vec{r}_j-\vec{k}_j) +2 O(\al^{-{1\over 2}})\ .
\ee
Remark that  $\vec{a}$ is fixed, once fixed $\Si_1$,...$\Si_{l-2}$.
We chose a coordinate system in which $\vec{r}_2=(1,0)$. Then, since
$\th_j=\angle
(\vec{r}_2,\vec{r}_j)$ satisfies  $|\th_j| = O(2^{-i})$ or $|\pi -\th_j|
= O(2^{-i})$  $\forall \; j\geq 2$ the $x$
and $y$ coordinates of every $\vec{k}_j$ $2\leq j\leq l$,
obey
\bqa
\vec{k}_j &=& \lp\pm [1+O(\al^{-{1\over 2}})] \cos O(2^{-i}),
 [1+O(\al^{-{1\over 2}})] \sin O(2^{-i})\rp\no\\
&=& \lp [\pm 1 + O(2^{-2i})] , O(2^{-i})\rp
\eqa
where we assumed $2^{-i}\geq\al^{-1/4}$ (otherwise all sectors are
automatically fixed).
On the other hand, the differences $\vec{k}_j-\vec{r}_j$ can be written
\bqa
&&\hskip-.8cm\vec{k}_j-\vec{r}_j = \vec{k}'_j-\vec{r}_j + 
O(\al^{-{1\over 2}})\no\\
&& \hskip.2cm = \lp \cos \lp \th_j+
O(\al^{-{1\over 4}})\rp, \sin \lp  \th_j+ O(\al^{-{1\over 4}})\rp
\rp - \lp \cos  \th_j, \sin  \th_j\rp +  O(\al^{-{1\over 2}})\no\\
&&\hskip.2cm =
\lp |\sin \th_j| O(\al^{-{1\over 4}}),
|\cos \th_j | O(\al^{-{1\over 4}})  \rp
+ O(\al^{-{1\over 2}})\ .
\eqa
For any $j\geq 2$ we know that
$|\sin \th_j | =  O(2^{-i})$ and $|\cos \th_j |=  O(1)$. For $j=1$,
since $k_{1}= -\sum_{j=2}^{l} k_{j}$ we can check that
$\max_{j\geq 2} \angle
(\vec{k}'_1, \vec{k}'_j) \leq
l O(2^{-i})$, hence $|\sin \th_1 | =  lO(2^{-i})$.
Therefore  we have
\bqa
\vec{k}_j-\vec{r}_j &=&
\lp  O(2^{-i} \al^{-{1\over 4}}),  O(\al^{-{1\over 4}})\rp
\quad \forall \; j>1\no\\
\vec{k}_1-\vec{r}_1 &=&
 l \lp O(2^{-i} \al^{-{1\over 4}}),  O(\al^{-{1\over 4}})  \rp\ .
\eqa
Inserting these results in the expressions for  $\vec{a}$ and $\vec{\vep}$
we have
\bqa
\vec{\vep} &=&
l \; O \lp 2^{-i} \al^{-{1\over 4}},  \al^{-{1\over 4}} \rp \no\\
\vec{a} &=& N (2,0) + O \lp 2^{-2i}, 2^{-i}\rp +
l\; O\lp 2^{-i}\al^{-{1\over 4}}, \al^{-{1\over 4}}\rp \no\\
 &=& N (2,0) + l \;  O \lp 2^{-2i}, 2^{-i}\rp
\eqa
where $N\in\{1,0,-1\}$.

\begin{figure}
\centerline{\psfig{figure=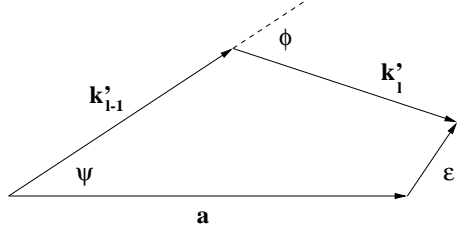,width=6cm}}
\caption{case 1, $|N|=1$}
\label{case1fig}
\end{figure}

Now we can bound $N_l$ in (\ref{bound}). We consider two cases.
First, let $|N|=1$. We rotate the coordinate system by
$\pi \de_{N,-1} + O(2^{-i})$ in such a way to make $\vec{a}$ run along the
positive $x$ axis (see Fig.\ref{case1fig}). In the new coordinate system the
coordinates of $\vec{\vep}$ obey, as before
\be
\vec{\vep} =
l\; O \lp 2^{-i} \al^{-{1\over 4}},  \al^{-{1\over 4}} \rp \ .
\ee
Remark that,
calling $\psi$ the angle $\angle (\vec{k}'_{l-1},\vec{a})$,
we must have   $\angle(\vec{k}'_{l},\vec{a})= \phi-\psi$.

\begin{figure}
\centerline{\psfig{figure=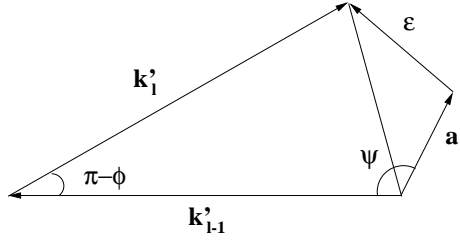,width=6cm}}
\caption{case 2, $|N|=0$}
\label{case2fig}
\end{figure}

Then the two components of the equation
\be
\vec{k}'_{l-1}+ \vec{k}'_{l} =
\lp \cos\psi, \sin\psi  \rp +
\lp \cos(\phi-\psi), \sin (\phi-\psi)  \rp = \vec{a} + \vec{\vep}
\ee
are
\be
\cos \psi + \cos (\phi-\psi) =
|\vec{a}| + l\; O(2^{-i}\al^{-{1\over 4}})\ \ ,\ \ 
\sin \psi - \sin (\phi-\psi) = l\; O(\al^{-{1\over 4}})\ .
\ee
The $y$ component implies that
\be
|2\psi-\phi| = l \; O(\al^{-{1\over 4}})\; ,\quad\quad
\psi = \frac{1}{2} \phi + l\; O(\al^{-{1\over 4}})
\ee
then $\phi$  is determinated  with precision
$O(\al^{-1/4})$ once $\psi$ has been fixed.
Remark this was not obvious
since   the maximal variation for $\phi$, without additional constraints,
is   $2^{-i}$ (remember $2^{-i}\le \phi\le 2^{-i+1}$). Therefore, for
$r_{l-1}$ fixed, $\th^s_l$  is restricted
to an interval   of width  $l\; O(\al^{-1/4})$.
Finally, we consider the $x$ component:
\bqa
\cos \psi + \cos \lp \phi-\psi\rp &=&
\cos \psi + \cos \psi \cos \lp \phi- 2 \psi\rp
- \sin \psi \sin\lp\phi- 2\psi\rp\no\\
&=& \left [ 2+ l^2 O(\al^{-{1\over 2}})\right ] \cos \psi + l O(2^{-i}
\al^{-{1\over 4}})\ .
\eqa
Then the angle $\psi$ is
\be
\psi = \cos^{-1} \lp \frac{|\vec{a}|}{2} \rp + l
O\lp {2^{-i}\al^{-{1\over 4}} \over 2^{-i} }\rp
\ee
and  $\th^s_{l-1}$  must be integrated
on an interval
of width  $l O(\al^{-1/4})$, instead of $\Om_{l-1}$.
This completes the proof for $|N|=1$.
Finally we consider the case $|N| =0$.  This time we rotate the coordinate
system by $O(2^{-i})$ or $\pi + O(2^{-i})$ so that $\vec{k}_{l-1}$
runs along the negative axis (see Fig.\ref{case2fig}).

Since $|\vec{a}|$ may be quite small we must perform a different
estimate. The angle $\phi$ is determined by
\be
\sin\lp  \frac{\pi-\phi}{2}\rp = \frac{|\vec{a}+\vec{\vep}|}{2}
= \frac{|\vec{a}|}{2} + l\; O(\al^{-{1\over 4}}).
\ee
Thus
\be
\phi = \pi - 2 \sin^{-1}\lp \frac{|\vec{a}|}{2} \rp + l\; O(\al^{-{1\over 4}})
\ee
and  $\th^s_l$  is restricted
to an interval   of width  $l\; O(\al^{-1/4})$, when $\vec{r}_{l-1}$
is held fixed.
To evaluate $\psi$ we apply the relation
\bqa
\left | \sin \lp \psi - \frac{\phi}{2} \rp \right |
&=& \frac{\left | \vec{\vep} \cdot \lp \cos\lp \frac{\pi-\phi}{2}\rp,
\sin \lp  \frac{\pi-\phi}{2}\rp \rp \right |}
{|\vec{a}|}\no\\
&\leq &  \frac{
l\; O(2^{-i}\al^{-{1\over 4}})}{O(2^{-i}) - l\;
O(\al^{-{1\over 4}})}
 \leq  l \; O(\al^{-{1\over 4}})\ ,
\eqa
where we applied the relation
\be
|\vec{a}+\vec{\vep}| = 2 \sin\lp \frac{\pi-\phi}{2}\rp\geq
 O(2^{-i})
\ee
that is proved with the hypothesis
$2^{-i}\leq \phi\leq 2^{-i+1}$. Then
\be
\psi = \frac{\phi}{2} + l\; O(\al^{-{1\over 4}})
\ee
hence $\th^s_l$  is restricted to an interval
of width $l\; O(\al^{-1/4})$.  This ends the proof.
\qed

\medskip
\noindent{\bf Acknowledgements}
\medskip

We thank C. Kopper and J. Magnen for many interesting discussions;
In particular the use of partial orderings (rather
than total orderings as in [DR1]) came from common work with C. Kopper.
We are especially grateful to M. Salmhofer: not only
his paper [S1] inspired this work, but he also explained
to us the meaning and physical importance of the uniform bounds on the 
derivatives of the self energy that we had not
included in a first version of this work.

\vskip.1cm

\medskip
\noindent{\large{\bf References}}
\medskip
\vskip.1cm
\noindent [AR1] A. Abdesselam and V.  Rivasseau, Trees, forests and jungles: a
botanical garden for cluster expansions, in Constructive Physics, ed by
V. Rivasseau, Lecture Notes in Physics 446, Springer Verlag, 1995.
\vskip.1cm

\noindent [AR2] A. Abdesselam and V. Rivasseau, Explicit Fermionic Cluster
Expansion, Lett. Math. Phys. {\bf 44} 1998 77-88.
\vskip.1cm

\noindent [BG] G. Benfatto and G. Gallavotti,
Perturbation theory of the Fermi surface in a quantum liquid.
A general quasi particle formalism and one dimensional systems, 
 Journ. Stat. Phys. {\bf 59} (1990) 541.
\vskip.1cm

\noindent [BGPS] G.Benfatto,  G.Gallavotti, A.Procacci, B.Scoppola,
Commun. Math. Phys. {\bf 160}, 93 (1994).

\noindent [BM] F.Bonetto, V.Mastropietro,
Commun. Math. Phys. {\bf 172}, 57 (1995).

\noindent [CR] C. de Calan and V. Rivasseau, Local existence of the Borel
transform in Euclidean $\phi^{4}_{4}$,  Commun. Math. Phys. {\bf 82}, 
69 (1981).
\vskip.1cm

\noindent [DR1] M. Disertori and V. Rivasseau,
Continuous Constructive Fermionic Renormalization, 
Annales Henri Poincar{\'e}, {\bf 1}, 1 (2000). 
\vskip.1cm

\noindent [DR2] M. Disertori and V. Rivasseau, Interacting Fermi liquid 
in two dimensions at finite temperature, Part II: Renormalization,
to appear.
\vskip.1cm

\noindent [FKLT] J. Feldman, H. Kn{\"o}rrer, D. Lehmann and E. Trubowitz, 
Fermi Liquids in Two Space Time Dimensions, in {\it
  Constructive Physics} ed. by V. Rivasseau, Springer Lectures Notes in
Physics, Vol 446, 1995.
\vskip.1cm

\noindent [FST] J. Feldman, M. Salmhofer and E. Trubowitz, Perturbation Theory 
around Non-nested Fermi Surfaces II.  Regularity of the Moving Fermi 
Surface, RPA Contributions, 
Comm. Pure. Appl. Math. {\bf 51} (1998) 1133;
Regularity of the Moving Fermi Surface, The Full Selfenergy,
to appear in Comm. Pure. Appl. Math. 
\vskip.1cm

\noindent [FT1]  J. Feldman and E. Trubowitz, 
Perturbation theory for Many Fermion Systems, Helv. Phys. Acta {\bf 63}
(1991) 156.
\vskip.1cm

\noindent [FT2]  J. Feldman and E. Trubowitz, The flow of an Electron-Phonon
System to the Superconducting State, Helv. Phys. Acta {\bf 64}
(1991) 213.
\vskip.1cm

\noindent [FMRT1] J. Feldman, J. Magnen, V. Rivasseau and E. Trubowitz,
An infinite Volume Expansion for Many Fermion Green's Functions,
Helv. Phys. Acta {\bf 65}
(1992) 679.
\vskip.1cm

\noindent [FMRT2] J. Feldman, J. Magnen, V. Rivasseau and E. Trubowitz,
An Intrinsic 1/N Expansion for Many Fermion System, Europhys. Letters 
{\bf 24}, 437 (1993).
\vskip.1cm

\noindent [FMRT3] J. Feldman, J. Magnen, V. Rivasseau and E. Trubowitz,
Ward Identities and a
Perturbative Analysis of a U(1) Goldstone Boson in a Many Fermion System, 
Helv. Phys. Acta {\bf 66}, 498 (1993).
\vskip.1cm

\noindent [G] M. Gevrey,  Sur la nature analytique des solutions des
 {\'e}quations aux d{\'e}riv{\'e}es partielles,
(Ann. Scient. Ec. Norm. Sup., 3 s{\'e}rie.
t. 35, p. 129-190) in {\it Oeuvres de Maurice Gevrey}
 pp 243 , ed. CNRS (1970).
\vskip.1cm

\noindent [L] A. Lesniewski, Effective Action for the Yukawa$_{2}$ 
Quantum Field Theory, Commun. Math. Phys. {\bf 108}, 437 (1987).
\vskip.1cm

\noindent [MR] J. Magnen and V. Rivasseau, A Single 
Scale Infinite Volume Expansion for
Three Dimensional Many Fermion Green's Functions,
Math.  Phys. Electronic  Journal, Volume 1,  1995.
\vskip.1cm

\noindent [R] V. Rivasseau, From perturbative to constructive renormalization,
Princeton University Press (1991).
\vskip.1cm

\noindent [S1] M. Salmhofer,
Continuous renormalization for Fermions and Fermi liquid theory,
Commun. Math. Phys.{\bf 194}, 249 (1998).

\noindent [S2] M. Salmhofer, Improved Power Counting and Fermi Surface
Renormalization, Rev. Math. Phys. {\bf 10}, 553 (1998).
\vskip.1cm


\end{document}